\documentclass[%
 reprint,
nofootinbib,
 amsmath,amssymb,
 aps,
 prd,
]{revtex4-1}

\usepackage{amsmath,amssymb} 
\usepackage{color}
\usepackage{graphicx}
\usepackage{subfig}          
\usepackage{hyperref}           
\usepackage{multirow,makecell}
\usepackage{braket,bm}
\usepackage{cancel}
\usepackage{leftidx}
\usepackage[text={17cm,24.5cm},centering]{geometry}
\usepackage{textcomp}
\usepackage{wasysym}
\usepackage{mathtools}
\usepackage{comment}
\usepackage{url}
\usepackage{caption}
\usepackage{orcidlink}
\usepackage{float}
\usepackage[strings]{underscore}

\makeatletter
\@addtoreset{equation}{section}
\makeatother

\def \be {\begin{equation}}
\def \ee {\end{equation}}
\def \ba {\begin{array}}
\def \ea {\end{array}}
\def \bea{\begin{eqnarray}}
\def \eea{\end{eqnarray}}

\def \G {\Gamma}

\def \D {\Delta}

\def \l {\lambda}

\def \r {\rho}
\def \o {\omega}
\def \O {\Omega}
\def \th {\theta}

\def \t {\tau}

\def \mbR {\mathbb R}
\def \mbS {\mathbb S}

\def \mF {\mathcal F}

\def \mK {\mathcal K}

\def \mR {\mathcal R}

\def \mT {\mathcal T}
\def \mU {\mathcal U}

\def \mc {\mathcal}

\def \inf {\infty}

\def\re#1{\textrm{Re}(#1)}
\def\im#1{\textrm{Im}(#1)}
\def \Re {{\textrm{Re}}}
\def \Im {{\textrm{Im}}}
\def \i {{\textrm{i}}}
\def \r {{\textrm{r}}}
\def \t {{\textrm{t}}}

\def \and {{\textrm{and}}}

\def\Order#1{{\cal O}\left(#1\right)}

\def\K{{\cal K}}

\def\R{{\cal R}}

\begin{document}

\title{Quasinormal mode/greybody factor correspondence for Kerr black holes}
\author{Zun-Xian Huang \orcidlink{0000-0002-0493-5562}}
\author{ Peng-Cheng Li \orcidlink{0000-0003-4977-2987}}
\email{pchli2021@scut.edu.cn}
\affiliation{
School of Physics and Optoelectronics, South China University of Technology, Guangzhou 510641, People’s Republic of China.
}
\date{\today}
\begin{abstract}	

\par We revisit the quasinormal-mode/greybody factor correspondence for Kerr black holes in the eikonal limit and develop a systematic WKB-based formulation by recasting the radial Teukolsky equation into a Schrödinger-type equation with a short-range potential. Building on earlier studies of the correspondence in rotating backgrounds, we extend the analysis to gravitational perturbations and incorporate higher-order WKB corrections beyond the leading eikonal approximation. For gravitational perturbations, the predicted greybody factors are in good agreement with numerical results obtained from the generalized Sasaki–Nakamura equation, with increasing accuracy at large angular quantum number. We also identify the breakdown of the correspondence in the superradiant regime, where the WKB assumptions cease to be valid. 
\end{abstract}
\maketitle
\pagenumbering{arabic}
\section{Introduction}
\label{Intro}
\par Quasinormal modes (QNMs) and greybody factors (GBFs) play central roles in characterizing the dynamical and radiative properties of black holes (BHs). QNMs describe the characteristic damped oscillations of perturbed BHs and dominate the gravitational-wave (GW) ringdown following compact binary mergers, providing powerful tests of the BH no-hair theorem \cite{Kokkotas:1999bd,Berti:2009kk,Konoplya:2011qq, Berti:2025hly}. GBFs, originating from the curvature-induced scattering of perturbations, determine the deviations of Hawking radiation from a perfect blackbody spectrum and also characterize the absorption and scattering properties of GWs by BHs~\cite{Hawking:1975vcx,PhysRevLett.85.5042}. Although these two quantities arise from seemingly different physical processes, recent work suggests that they may be linked in a nontrivial and potentially universal manner  \cite{Oshita:2023cjz,Okabayashi:2024qbz,Rosato:2025ulx}. 

A key development in this direction is the recent discovery by Oshita and collaborators \cite{Oshita:2023cjz,Okabayashi:2024qbz} (and also \cite{Nair:2025anr}) that the high-frequency portion of the ringdown spectrum encodes GBFs of the BH. Specifically, for the  $(l,m)=(2,2)$ mode, the GW spectral amplitude satisfies
\begin{align}
|\tilde{h}_{l m} (\omega)| \simeq c_{l m} \frac{\sqrt{1-\Gamma_{l m} (\omega)}}{\omega^3}, \quad\text{for}\hspace{2pt}  \ \omega \gtrsim f_{l m},
\end{align}
where $c_{l m}$ is a constant corresponding to the GW amplitude, $\Gamma_{l m} (\omega)$ is the GBF and $f_{l m}$ denotes the  real part of the spectrum of the fundamental QNMs. Moreover, GBFs were shown to be stable under small spacetime deformations and related to quantities reconstructible from spectrally unstable QNMs \cite{Kyutoku:2022gbr,Rosato:2024arw,Oshita:2024fzf,Xie:2025jbr}, suggesting a deep interplay between Hawking radiation and dynamical GW observables.

Motivated by these insights, Konoplya and Zhidenko \cite{Konoplya:2024lir} established a rigorous QNM/GBF correspondence for spherically symmetric and asymptotically flat (or de Sitter) BHs using WKB techniques in the eikonal approximation \cite{Iyer:1986np,Konoplya:2019hlu}. 
Their work demonstrated that GBFs can be directly reconstructed from the QNM spectrum. This correspondence has since been explored in a variety of BH backgrounds, including rotating BHs \cite{Konoplya:2024vuj,Pedrotti:2025idg}, Schwarzschild-de Sitter BHs \cite{Malik:2024cgb}, regular BHs \cite{Malik:2025erb,Heidari:2024bbd,Skvortsova:2024msa,Dubinsky:2025wns,Tang:2025mkk,Lutfuoglu:2025ohb,Shi:2025gst,Bolokhov:2025lnt,Malik:2025dxn,Dubinsky:2025nxv,Lutfuoglu:2025blw,Malik:2025qnr}, BHs in higher dimensions \cite{Dubinsky:2025ypj,Han:2025cal}, BHs in modified  gravities \cite{Dubinsky:2024vbn,AraujoFilho:2025hkm,Heidari:2025oop,Lutfuoglu:2025ldc}, BHs surrounded astrophysical environments \cite{Hamil:2025pte,Yan:2025pvp,Konoplya:2025mvj,Sajadi:2025kah}, and wormholes \cite{Malik:2024wvs,Bolokhov:2024otn}, highlighting its theoretical robustness and broad applicability. In particular, previous studies have provided evidence for the correspondence in rotating spacetimes for scalar and vector perturbations  \cite{Konoplya:2024vuj,Pedrotti:2025idg}.

Despite this progress, a systematic Kerr treatment based directly on the radial Teukolsky equation \cite{Teukolsky:1972my} and its associated short-range Schrödinger-type form remains worthwhile, especially if one wishes to clarify the underlying WKB structure and assess the role of higher-order corrections in a unified way. Moreover, the gravitational perturbation case is of particular physical interest because of its direct relevance to BH spectroscopy and GW observations.

In this work, we develop such a WKB-based formulation  for Kerr BHs in the eikonal limit. By recasting the radial Teukolsky equation into a Schr\"odinger-type form with a short-range potential, we derive connection formulas relating Kerr QNMs and GBFs, extend the analysis beyond the leading eikonal order, and test the correspondence numerically for gravitational perturbations. In this sense, the main new ingredient of the present work is the systematic Kerr derivation together with its application to the gravitational sector.

To test the correspondence, we compute Kerr QNMs and GBFs with high numerical precision.
The QNM frequencies for gravitational perturbations ($s=-2$) are obtained using two independent state-of-the-art solvers, while the GBFs are computed by solving the Sasaki-Nakamura equation. We find excellent agreement between the correspondence predictions and the exact GBFs: deviations remain below $10^{-2}$ in nonsuperradiant regimes and decrease rapidly with increasing $l$, consistent with expectations from WKB theory. Additionally, we explicitly identify the breakdown of the correspondence in the superradiant regime, where GBFs become negative and the WKB-based derivation is no longer applicable.

\par This paper is organized as follows. In Sec.~\ref{theofram}, we review the Teukolsky equations and their variants, discuss the application of the WKB method to Kerr BHs, and derive the QNM/GBF correspondence in the eikonal limit, including higher-order corrections. Section~\ref{Numexam} presents high-accuracy methods for computing QNMs and GBFs, and compares the results obtained from the correspondence with exact numerical calculations. We conclude in Sec.~\ref{conclusion} with a summary of our findings and possible directions for future work.
Throughout this paper, we use geometric units $c = G =1$.
\section{Theoretical framework}
\label{theofram}
\par  In the Boyer-Lindquist coordinate system $(t,r, \theta,\phi)$, the line element of the Kerr metric is given by 
\cite{Boyer:1966qh}

\begin{align}
\label{Kerrmetric}
    ds^2 = &- \left( 1- \frac{2 M r}{\Sigma}\right)dt^2 - \frac{4 M  r a \sin^2 \theta }{\Sigma}dt d\phi+\frac{\Sigma}{\Delta}dr^2         \nonumber\\
           & + \Sigma d\theta^2  + \left(r^2 + a^2 + \frac{2 M  r a^2 \sin^2 \theta}{\Sigma}\right) \sin^2 \theta d\phi^2,
\end{align}
with
\be
        \Sigma\equiv r^2+ a^2 \cos^2 \theta   ,\quad \Delta\equiv r^2- 2 M r + a^2 ,
\ee
where $a\equiv J/M$ is the spin parameter, and we assume $0\leq a\leq M$. The zeros of $\Delta$, $r_{\pm} \equiv M \pm\sqrt{M^ 2 - a^2}$, correspond to the event horizon and the inner horizon of the BH. When $a=0$, it reduces to the Schwarzschild metric.
\subsection{Perturbation equations}

To study perturbations of the Kerr spacetime, one may linearize Einstein’s equations around the Kerr background. However, directly perturbing the metric leads to a complicated set of coupled partial differential equations that are difficult to solve. 
Teukolsky \cite{Teukolsky:1972my} instead focused on perturbations of certain Newman–Penrose components of the curvature, specifically the Weyl scalars $\Psi_0$ and $\Psi_4$, which encode the ingoing and outgoing radiative degrees of freedom. Remarkably, these curvature perturbations satisfy a single, decoupled master equation valid for fields of arbitrary spin weight. This Teukolsky equation admits separation of variables, with solutions decomposing into spin-weighted spheroidal harmonics and radial functions, greatly simplifying the analysis of BH perturbations. To be more specific, the perturbation fields $\Psi_{(s)}$  have the following decomposition
\be
\label{separationofvariables}
\Psi_{(s)}(t,r,\th,\phi)= R_{slm}(r) S_{slm}(\theta,\phi)     e^{-i \o t},
\ee
in which $s$ is the spin weight of the field, i.e. scalars ($s=0$), spinors ($s=\pm1/2$), vectors ($s=\pm1$) and tensor perturbations ($s=\pm2$), $l$ is the spheroidal harmonic index and $m$ is the azimuthal harmonic index with $-l\leq m\leq l$. For simplicity, we omit the indices $s,l,m$ in the following. Then the single Teukolsky equation can be separated into radial and angular equations. The radial equation for $R(r)$ is given by \cite{Teukolsky:1973ha}
\begin{align}
\label{radialTeukolskyequation}
\Delta^{-s} \dfrac{d}{dr}\left( \Delta^{s+1} \dfrac{dR}{dr} \right)-V_T(r)R = 0,
\end{align}
with
\be
V_T(r)= \l-4is\omega r - \dfrac{K^2 - 2is(r-M)K}{\Delta},
\ee
where $K\equiv (r^2+a^2)\o-ma$ and $\l$ is the separation constant related to the angular equation, which is \cite{Teukolsky:1973ha}
\begin{align}
&\Big[\frac{1}{\sin\th}\frac{d}{d\th}\Big(\sin\th \frac{d}{d\th} \Big)-a^2\omega^2\sin^2\th-\frac{(m+s\cos\th)^2}{\sin^2\th}\nonumber\\
&-2a\omega s\cos\th+s+2ma\omega+\l\Big]S=0. 
\label{angularTeukolskyequation}
\end{align}
\par The radial  equation (\ref{radialTeukolskyequation}) has the following asymptotic solutions 
\be
R \sim \left\{ \begin{array}{ll}
Z_\i^{\mathrm{T}} \frac{e^{- i\o r_* }}{r}   +Z_\r ^{\mathrm{T}}\frac{e^{ i\o  r_* }}{r^{2s+1}}  , \quad \quad &     r \rightarrow + \infty , \\ \\
Z_\t^{\mathrm{T}} \Delta^{-s} e^{ - i p r_* },  \quad \quad &  r \rightarrow  r_+,
\end{array} \right.
\label{ras} 
\ee
where $p=\o-m\O_h$ and $\O_h=a/(2Mr_+)$ is the angular velocity of the BH horizon, and $r_*$ is the tortoise coordinate defined by
\be
\frac{dr_*}{dr}=\frac{r^2+a^2}{\Delta}.
\ee
These boundary conditions describe a standard scattering process in which  an incident wave from spatial infinity gives rise to a reflected wave  and a transmitted wave. For $p<0$, namely $\omega<m\Omega_h$, the scattering wave is superradiantly amplified. This can be explained from the boundary condition at the horizon. When superradiance is triggered, the wave seen by an observer at infinity is outgoing, while being ingoing by an local observer, which means the energy flows out of the BH and the corresponding scattering wave is amplified.

A notable feature of the radial equation \eqref{radialTeukolskyequation} is that, when rewritten in a Schr\"odinger-like form, which is given by \cite{Teukolsky:1973ha}
\be\label{TeukolskyradialY}
\frac{d^2Y}{dr_*^2}+(\omega^2-V_Y)Y=0,
\ee
where $Y$ is a new radial function transformed from the radial function $R$ and $V_Y$ is the potential related to $V_T$,
the potential $V_Y$ is long-ranged, as opposed to a short-ranged potential, which falls off faster than $1/r_*$ at spatial infinity $r_*\to \infty$ or horizon $r_*\to -\infty$.\footnote{There are two except cases. One is the scalar perturbation, which corresponds to $s=0$. The other is taking the eikonal limit $l\to \infty$ \cite{Yang:2012he}. In these two cases, the potential $V_Y$ is short-ranged. } This feature makes the solutions of the radial Teukolsky equation exhibit a more complicated asymptotic structure than that of the Regge-Wheeler \cite{Regge:1957td} or Zerili \cite{Zerilli:1970se} equation, since the latter has a short-ranged potential. For example, as shown in \eqref{ras}, for gravitational perturbations, the incident and outgoing waves have different power-law dependence of $r$ at infinity, the former  has a higher power of $r$  in its asymptotic amplitude, which will overwhelm the other solution. Therefore, it is challenging to solve the radial Teukolsky equation accurately. This difficulty hinders the usage of the standard WKB method to study QNMs and GBFs in a unified manner \cite{Iyer:1986np}, as this method is suitable for the perturbation equation of the form that owns a short-ranged potential.\footnote{In a recent paper \cite{Tang:2025qaq}, the authors obtained the QNMs of Kerr BHs by directly applying the WKB method to the radial Teukolsky equation \eqref{TeukolskyradialY}. Their results exhibit better agreement with numerical calculations than earlier WKB approaches based on equations with short-ranged potentials. It is worth noting, however, that in those earlier works the separation constant was typically evaluated using a slow-rotation expansion \cite{Seidel:1989bp}, which may introduce additional systematic errors at moderate and high spins.} 

In recent years, two main schemes have been proposed to transform the long-ranged potential of the radial Teukolsky equation to a short-ranged one. The first one is proposed by Chandrasekhar and Detweiler in a series of papers \cite{Chandrasekhar:1975zz,10.1098/rspa.1976.0022,Chandrasekhar:1976zz,Chandrasekhar:1977kf} (see \cite{TorresdelCastillo:1992zq,Arbey:2019mbc} for further extensions), which  successfully transformed the radial Teukolsky equation into the Schr\"odinger-like form with  short-range potentials for perturbations of various fields. They found necessary to define a new tortoise  coordinate $d\tilde{r}_*/dr=\rho^2/\Delta$, where $\rho^2=r^2+a^2-am/\omega$. However, as pointed out in \cite{Arbey:2025dnc}, this formalism is inconvenient to apply directly, since $\tilde{r}_*(r)$ may not be a monotonic function of $r$ in the superradiant regime $\omega<m\Omega_h$. To overcome this issue, some adjustment needs to be introduced. 

Alternatively, Sasaki and  Nakamura \cite{Sasaki:1981kj,Sasaki:1981sx,Sasaki:2003xr} have found a class of transformations which convert the radial Teukolsky equation for gravitational perturbations $s=-2$ into a form which has a short-ranged potential and reduces to the Regge-Wheeler equation in the $a\to 0$ limit, but does not have the Schr\"odinger-like form. The SN transformations were then generalized for arbitrary integer spin weight $s$ in \cite{Hughes:2000pf} (and \cite{Lo:2023fvv} for more details). The generalized SN (GSN) equation can be written as 
\begin{equation}
\label{GSNeqn}
    \left[\dfrac{d^2 }{d r_*^2} - \mF(r)\dfrac{d}{dr_*} - \mU(r)\right]X = 0,
\end{equation}
where the explicit forms of the  potentials $\mF(r)$ and $\mU(r)$ can be found in Refs.~\cite{Sasaki:1981sx,Lo:2023fvv}. This equation has been extensively used in the study of linear dynamics of Kerr BHs, particularly the calculations of GWs from extreme-mass-ratio inspirals around Kerr BHs \cite{Pound:2021qin}. We can further perform a replacement 
\be
X(r_*)=\exp\left(\frac12\int^{r_*}\mF(x)dx\right)\psi(r_*),
\ee
to eliminate first-order derivative term, and obtain 
\be\label{standardEq}
\frac{d^2\psi}{dr_*^2}+Q(r)\psi=0,
\ee
where
\be
Q(r)=-\mU(r)+\frac12 \mF'(r_*)-\frac14 \mF(r)^2.
\ee
One can check that, $Q\to\omega^2+\mc O(r_*^{-2})$ as $r_*\to\infty$ and $Q\to\left(\omega-m\Omega_h\right)^2+\mc O(r_*^{-2})$ as $r_*\to-\infty$. The above equation is indeed of the Schr\"odinger-like form and has a short-ranged potential.
\subsection{WKB method for Kerr BHs}
\label{wkbinkerrbh}

\par In this subsection we give a brief review of the basic idea about the WKB method and present the important formulae (see \cite{Iyer:1986np} for more details).  Our starting point is the standard  Schr\"odinger-like equation \eqref{standardEq}, for which we introduce 
\be\label{potential}
Q(r)=\omega^2-V(r_*,\omega),
\ee
to make the analogy with the Schwarzschild BH more direct. Here the $V(r_*,\omega)$ is the scattering potential, which is short-ranged and may also depend on the frequency of the perturbation. 

\begin{figure}\centering
\includegraphics[scale=0.4]{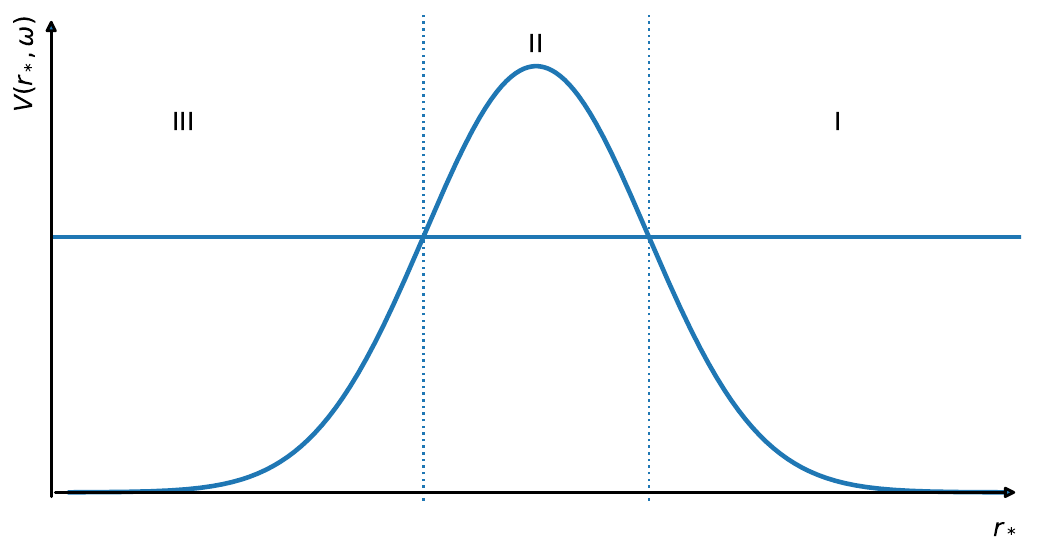}
\caption{Different regions for the radial wave function
. The WKB analysis is valid in regions I and III. The near-peak region is located at II.}
\label{wkbregion}
\end{figure}
The potential has the general shape shown in Fig.~\ref{wkbregion}. For a wave incident on the barrier from region I, one can use the WKB method to calculate the amplitude of the waves transmitted to region III and reflected back to region I. 
When $r_* \rightarrow \inf$, $V \rightarrow 0$, thus there are two possible asymptotic solutions at spatial infinity, one is  $\psi\rightarrow e^{+i \omega r_*}$ which denotes an outgoing wave  and the other is $\psi\rightarrow e^{-i \omega r_*}$ which denotes an incoming wave from infinity. 
Similarly, for $r_* \rightarrow -\inf$, we have $\psi\rightarrow e^{+ip r_*}$ for a wave coming from the event horizon, while $\psi\rightarrow e^{-ip r_*}$ is a wave going toward the event horizon. We denote these four WKB solutions  by the notation
\begin{align}
\psi^\text{I}_{+}\sim e^{+i \omega r_*},\quad \psi^\text{I}_{-}\sim e^{-i \omega r_*} , \quad  &\text{region I},\\
\psi^\text{III}_{+} \sim e^{+i p r_*},\quad \psi^\text{III}_{-}\sim e^{-i p r_*} , \quad  &\text{region III}.
\end{align}
Then the general solutions  in regions I and III can be written as 
\begin{align}
\psi^\text{I}\sim Z^\text{I}_\text{i} \psi^\text{I}_{-} + Z^\text{I}_\text{o} \psi^\text{I}_{+},\qquad  &\text{region I} , \\ 
\psi^\text{III}\sim Z^\text{III}_\text{i}  \psi^\text{III}_{+} + Z^\text{III}_\text{o} \psi^\text{III}_{-},\qquad   &\text{region III} ,
\label{gsuriv}
\end{align}
where we have introduced four complex amplitudes $Z^\text{I}_\text{i}$, $Z^\text{I}_\text{o}$, $ Z^\text{III}_\text{i}$ and $Z^\text{III}_\text{o}$, respectively for ingoing and outgoing waves in  regions I and III. We connect the amplitudes in region I with those in III via the linear matrix $\mbS$
\begin{equation}
\label{lms}
\begin{pmatrix}
Z_\text{o}^{\text{III}} \\
Z_\text{i}^{\text{III}}
\end{pmatrix}
 =
\begin{pmatrix}
\mbS_{11} & \mbS_{12} \\
\mbS_{21} & \mbS_{22}
\end{pmatrix}
\begin{pmatrix}
Z_\text{o}^{\text{I}} \\
Z_\text{i}^{\text{I}}
\end{pmatrix}.
\end{equation}
To determine the elements of the scattering matrix $\mbS$, we match the WKB solutions in regions~I and~III with the solution in region~II across the two classical turning points defined by $Q(r)=0$. In order to incorporate QNMs into the scattering formalism, we focus on the regime in which $\omega^2$ lies close to the maximum of the effective potential $V(r_*)$, located at $r_{*0}$. In this case, the two classical turning points are situated in close proximity. QNMs correspond to purely outgoing conditions (for the potential), such that no incident wave is present and the reflected and transmitted amplitudes become comparable in magnitude. Near the peak of the potential, the effective potential is expanded as a Taylor series about $r_{*0}$, then the wave equation in region~II can be solved in terms of parabolic cylinder functions. By matching this solution to the WKB solutions in regions~I and~III across the turning points, one obtains the scattering matrix $\mbS$, from which both the GBFs and the QNM condition can be derived. The expression of $\mbS$ is given by
\begin{equation}
\label{cfresult}
\mbS= \left(\begin{array}{cc} e^{i \pi \nu} & \frac{i \mbR^2  e^{i \pi \nu} \sqrt{2 \pi}}{\Gamma(\nu +1)} \\
\frac{\mbR^{-2} \sqrt{2 \pi}}{\Gamma(-\nu)} & - e^{i \pi \nu}  \end{array}\right),
\end{equation}
where $\mbR$ is a function of $\nu$ and its explicit form is given in \cite{Iyer:1986np}.

Strictly speaking, in the Kerr case the effective potential \eqref{potential} entering the Schr\"odinger-type radial equation after the GSN transformation may be frequency dependent and generally complex-valued. Therefore, the phrase ``single-peaked potential" should not be interpreted literally as the maximum of an ordered real-valued function. What is actually required by the WKB derivation is the existence of the standard single-barrier configuration underlying the matching procedure, namely a dominant barrier region together with the associated turning-point/barrier-top structure. For the parameter range considered in this work, we find that in the nonsuperradiant regime the real part of the transformed effective potential exhibits a clear single-barrier profile, while the imaginary part remains subleading for the nonextremal cases studied here. This is also consistent with the eikonal scaling, in which the leading barrier structure is controlled by the dominant large-$l$ terms.

When considering an ordinary scattering problem, the physical boundary conditions have the following form
\be
\psi \sim \left\{ \begin{array}{ll} 
e^{ - i \Omega r_* } +\R e^{ i \Omega r_* },  \quad \quad &  r_* \rightarrow + \infty , \\ 
\mT  e^{- i(\Omega-m\Omega_h)  r_* }, \quad \quad &     r_* \rightarrow - \infty,
\end{array} \right.
\ee
where the real frequency of the wave is denoted by $\Omega$.\footnote{In what follows, we use $\Omega$ for the real frequency variable in the scattering/greybody-factor problem, and $\omega$ for the generally complex QNM frequency, in order to avoid confusion between the two.} Subject to the fact that $Q(r)$ has a dominant real part in the eikonal limit and the unitarity of the $\mbS$ matrix, the amplitudes satisfy the conservation equation 
\be
|\mT|^2+|\mR|^2 = 1.
\ee
The greybody factor, defined as the transmission probability is given by \cite{Iyer:1986np}
\begin{align}\label{GBFformula}
\G_{lm}^s(\Omega)=|\mT|^2=\frac{1}{1+e^{2\pi i \K}},
\end{align}
where we have introduced $\K=\nu+\frac{1}{2}$. To determine $\K$ and so the GBF, we need to solve the following equation \cite{Konoplya:2019hlu}
\begin{align}
    \Omega^2=&V_{0}+A_2(\K^2)+A_4(\K^2)+\ldots \nonumber\\
             &-i\K\sqrt{-2V_{2}}\left(1+A_3(\K^2)+A_5(\K^2)\ldots\right),
\label{gWKBe}
\end{align}
where $A_\i$ is the $i$th order WKB correction term beyond the eikonal approximation and depends on $\K$ and the derivatives of the  radial potential at its maximum up to the order 2i, and $V_{\i}$ denotes the $i$th derivative of $V$ with respect to $r_*$ at the point $r_{*0}$. Then the solution  can be formally written as
\be
\label{GBFQNMcondi}
-i\K=\frac{\Omega^2- V_{0}}{\sqrt{-2V_{2}} }+\sum_{\i=2}^k\Lambda _\i(\K),
\ee
here $\Lambda_\i$ is also the $i$th order WKB term and its explicit form up to $k=13$  can be found in  Refs.~\cite{Iyer:1986np,Konoplya:2003ii,Matyjasek:2017psv}. It is worth noting that, because $i\mK$ is a real function of the frequency $\Omega$, the GBF given by Eq.~\eqref{GBFformula} is manifestly non-negative. As a result, the WKB approximation is unable to describe the scattering problem in the superradiant regime, where the GBF is expected to be negative. This point has been verified for the scattering of a vector wave by Kerr BHs \cite{Pedrotti:2025idg}.

On the other hand, QNMs may be viewed as a special scattering problem in which no incoming wave is allowed from either asymptotic end. More precisely, one imposes a purely outgoing-wave condition at spatial infinity and a purely ingoing-wave condition at the horizon. In terms of the amplitudes introduced above, this means
$Z_\text{i}^\text{I}=0$ and $Z_\text{i}^\text{III}=0$.
Unlike the ordinary scattering problem, this is therefore a homogeneous boundary-value problem. A nontrivial QNM solution exists only if the connection formula (\ref{cfresult}) is compatible with these two conditions without forcing all amplitudes to vanish identically. From Eq. (\ref{cfresult}), the coefficient relating the forbidden incoming component to the allowed wave amplitudes contains the factor  $1/\Gamma(-\nu)$. Since the gamma function has no zeros, this coefficient can vanish only when $\Gamma(-\nu)$ develops a pole. This happens precisely when $-\nu=0,-1,-2,...$, namely  $\nu=n$, $n=0,1,2,\dots$. This is the usual WKB quantization condition for quasinormal modes. Therefore, Eq. \eqref{GBFQNMcondi} can be used to determine the complex QNM frequencies by replacing 
$\Omega\to\omega$ and $\mK\to n+\frac12$, where $\omega$ is  complex and $n$ is the overtone number.

To summarize, by recasting the radial Teukolsky equation into the Schr\"odinger-like form \eqref{standardEq} with a short-range potential, the WKB method---previously successfully applied to the calculation of QNMs and GBFs for spherically symmetric BHs such as the Schwarzschild BH---can also be applied to Kerr BHs. In particular, all WKB-based formulas developed for spherically symmetric cases remain valid for Kerr BHs, except that the separation constant appearing in the effective potential \eqref{standardEq} depends on the frequency $\omega$ and must be determined from the angular perturbation equation.

\subsection{QNM/GBF correspondence for Kerr BHs}
\par Since both the GBFs and QNMs are intimately connected to the effective potential of the radial perturbation equation, and are governed by equations of the same form within the WKB framework, it is natural to expect a correspondence between these two quantities. Indeed, Konoplya and Zhidenko \cite{Konoplya:2024lir} established a correspondence between GBFs and QNMs for spherically symmetric BHs in the eikonal approximation. The central idea of their derivation is that the WKB formula \eqref{gWKBe} can be employed to compute both QNMs and GBFs. Consequently, the QNM frequencies can be used to extract information about the effective potential, which can in turn be utilized to determine the GBFs. In the eikonal approximation, this procedure can be carried out fully analytically. Moreover, the derivation of Ref.~\cite{Konoplya:2024lir} can be straightforwardly extended to Kerr BHs, since the WKB method remains applicable in the rotating case and their analysis does not rely on the specific functional form of the effective potential.

Taking the eikonal approximation, the effective potential can be expanded as \cite{Konoplya:2023moy}
\be
\label{vlargel}
V(r_*,\o)=l^2U_0+lU_1+U_2+l^{-1}U_3+\cdots,
\ee
where $U_\i$ may be functions of $r_*$ and $\omega$. At the leading order of the eikonal approximation, the QNM formula \eqref{GBFQNMcondi} can be truncated at the first order of the WKB expansion, so one has
\be
 n+\frac{1}{2}=  \frac{i(\o^2-l^2U_{00} )}{l \sqrt{-2 U_{02}}}+\mc O(l^{-1}).
\ee
For the fundamental QNM, $n=0$, the solution to the above equation is 
\be
\omega_0=l \sqrt{U_{00}}-\frac{i}{2}\sqrt{\frac{-U_{02}}{2U_{00}}}+\mc O(l^{-1}),
\ee
from which we can express the effective potential and its derivative in terms of the frequency  of the fundamental QNM, 
\bea
U_{00}&=&l^{-2}\Re(\omega_0)^2+\mc O(l^{-1}),\\
U_{02}&=&8l^{-2}\Re(\omega_0)^2\Im(\omega_0)^2+\mc O(l^{-1}).
\eea
\par  Similarly, the first-order WKB formula for GBFs \eqref{GBFQNMcondi} in the eikonal limit  $l\to \infty$ is given by
\begin{eqnarray}\label{eikonal-K}
-i\K&=&\frac{\Omega^2- l^2U_{00}}{l\sqrt{-2U_{02}} }+\mc O(l^{-1}).
\end{eqnarray}
Substituting this into \eqref{GBFformula}, one can derive the expression for the GBF
\be
\G_{lm}^s(\Omega)=\left(1+e^{2\pi\dfrac{\Omega^2-\Re(\omega_0)^2}{4\Re(\omega_0)\Im(\omega_0)}}\right)^{-1}+\mc O(l^{-1}),
\ee
which, as expected, has the same form with the result for Schwarzschild BHs  \cite{Konoplya:2024lir}. This correspondence is exact only in the eikonal limit, and extending it beyond the eikonal regime requires the inclusion of higher-order corrections to \eqref{eikonal-K}. As demonstrated in \cite{Konoplya:2024lir}, all the corrections up to $\Order{l^{-2}}$ can be derived from six-order WKB formula and depend only on the frequencies of the fundamental mode $\omega _0 $ and the first overtone $\omega _1$. The result can be written as 
\begin{equation}
\label{6thfm}
-i\K = -\frac{\Omega^2-\re{\omega_0}^2}{4\re{\omega_0}\im{\omega_0}}+\Delta_1+\Delta_2+\Delta_f+\Order{l^{-3}},
\end{equation}
where
\begin{align}
\Delta_1=\frac{\re{\omega_0}-\re{\omega_1}}{16\im{\omega_0}}+\Order{l^{-2}},
\end{align}
\begin{align}
\Delta_2=
&-\frac{\Omega^2-\re{\omega_0}^2}{32\re{\omega_0}\im{\omega_0}}
\Big(\frac{(\re{\omega_0}-\re{\omega_1})^2}{4\im{\omega_0}^2}  \nonumber\\ 
&-\frac{3\im{\omega_0}-\im{\omega_1}}{3\im{\omega_0}}\Big)
+\frac{(\Omega^2-\re{\omega_0}^2)^2}{16\re{\omega_0}^3\im{\omega_0}}  \nonumber\\
&\times \left(1+\frac{\re{\omega_0}(\re{\omega_0}-\re{\omega_1})}{4\im{\omega_0}^2}\right)+\Order{l^{-3}},
\end{align}
and
{\small
\begin{align}\label{delta3}
&\Delta_f=\frac{(-\Omega^2+\re{\omega_0}^2)^3}{32\re{\omega_0}^5\im{\omega_0}}\Big(1+\frac{\re{\omega_0}(\re{\omega_0}-\re{\omega_1})}{4\im{\omega_0}^2}\nonumber\\
&+\re{\omega_0}^2\big(\frac{(\re{\omega_0}-\re{\omega_1})^2}{16\im{\omega_0}^4}-\frac{3\im{\omega_0}-\im{\omega_1}}{12\im{\omega_0}}\big)\Big)  \nonumber\\&+\Order{l^{-3}}.
\end{align}
}

We emphasize that the derivation of the higher-order corrections in \cite{Konoplya:2024lir} relies only on the generic WKB expansion of a one-dimensional Schr\"odinger-type equation with a single potential barrier, and does not assume any specific functional form of the effective potential. Since the Kerr perturbation equation, after the GSN transformation, falls into the same class of one-dimensional Schrödinger-type equations with a short-ranged effective potential and the standard single-barrier structure required by the WKB matching procedure. Both the QNM frequencies and the GBF parameter $\K$ are governed by the same WKB equation \eqref{GBFQNMcondi}, the only difference being which quantity is regarded as input. In the eikonal regime, after expanding the effective potential as in Eq. \eqref{vlargel}, one may derive analytic expressions for $\omega_0$, $\omega_1$ and $\K$ in terms of the same set of expansion coefficients and their derivatives. Eliminating these coefficients in favor of $\omega_0$ and $\omega_1$ then yields Eqs. \eqref{6thfm}-\eqref{delta3}. Therefore, the final form of these relations does not depend on the specific functional form of the Kerr effective potential, provided that the WKB equation \eqref{GBFQNMcondi} and the eikonal expansion \eqref{vlargel} are valid. Nevertheless, the validity of this extension is ultimately justified {\em a posteriori} by the excellent agreement with exact numerical GBFs \cite{Konoplya:2024vuj,Pedrotti:2025idg}.

\section{Numerical examination}
\label{Numexam}
In this section, we assess the accuracy and the range of applicability of the QNM/GBF correspondence \eqref{6thfm} for Kerr BHs. Our strategy is as follows. We first use publicly available codes to obtain the exact values of $\omega _{0} $ and $\omega _{1}$, which are then substituted into Eqs.~\eqref{6thfm} and \eqref{GBFformula} to compute the GBFs via the QNM/GBF correspondence. Independently, we numerically solve the GSN equation \eqref{GSNeqn} using public codes to obtain the exact values of the GBFs. By comparing the GBFs obtained from these two methods, we can examine the performance of the QNM/GBF correspondence over a wide range of parameters, including $a$, $l$ and $m$. Since the QNM/GBF correspondence for scalar and vector perturbations has already been verified in a similar manner \cite{Konoplya:2024vuj,Pedrotti:2025idg}, here we focus exclusively on gravitational perturbations with $s=-2$.
\subsection{High-accuracy calculation of QNMs}
\begin{table}
\centering
\caption{The frequencies of the fundamental mode and the first overtone mode  of gravitational perturbations for $a=0.3,0.6,0.9$ with $l=2,m=0$. }
\label{qnm.3.6.9}
\begin{tabular}{|c|c|c|} 
\hline
$a$ & $\omega _0 $     & $\omega _1 $                     \\ 
\hline
0.3      & 0.37698506 - 0.08835328i       & 0.35111569 - 0.27182132i    \\ 
\hline
0.6       &     0.38805392 - 0.08599467i   &  0.36553354 - 0.26375359i \\ 
\hline
0.9        &   0.41200447 - 0.07848270i    &  0.39347288 - 0.23847955i \\ 
\hline
\end{tabular}
\end{table}

\begin{table}
\centering
\caption{The  $\o_0$ and $\o_1$  of gravitational perturbations for $a=0.25,0.50,0.75,0.99$ with $l=m=2$. }
\label{qnm.25.50.75.99}
\begin{tabular}{|c|c|c|} 
\hline
$a$ & $\omega _0 $     & $\omega _1 $                     \\ 
\hline
0.25      &    0.41051791 - 0.08804765i   & 0.38835284 - 0.26938268i   \\ 
\hline
0.50       &  0.46412303 - 0.08563883i     &  0.44740704 - 0.26022455i \\ 
\hline
0.75        &  0.55681739 - 0.07860254i   & 0.54696353 - 0.23735737i \\ 
\hline
0.99        &0.87089266 - 0.02939042i    & 0.87064516 - 0.08817538i \\ 
\hline
\end{tabular}
\end{table}

\begin{table}
\centering
\caption{The $\o_0$ and $\o_1$  of gravitational perturbations for $m=-2,-1,0,1,2$ with $a=0.8,l=2$. }
\label{qnm-2-1012}
\begin{tabular}{|c|c|c|} 
\hline
$m$ & $\omega _0 $     & $\omega _1 $                     \\ 
\hline
-2      & 0.30331342 - 0.08851224i      & 0.26427308 - 0.27722963i    \\ 
\hline
-1      &   0.34535578 - 0.08600346i    & 0.31727276 - 0.26576826i  \\ 
\hline
0       &   0.40191735 - 0.08215627i    &   0.38256890 - 0.25073618i\\ 
\hline
1       &  0.48023071 - 0.07795498i     &  0.46761193 - 0.23576610i  \\ 
\hline
2       &     0.58601697 - 0.07562955i  &   0.57792240 - 0.22814894i \\ 
\hline
\end{tabular}
\end{table}

\begin{table}
\centering
\caption{The $\o_0$ and $\o_1$  of gravitational perturbations for $l=m=2,3,4,5$, where $a=0.9$. }
\label{qnm2345}
\begin{tabular}{|c|c|c|} 
\hline
$l,m$ & $\omega _0 $     & $\omega _1 $                     \\ 
\hline
2      &  0.67161427 - 0.06486924i  &  0.66765755 - 0.19525207i  \\ 
\hline
3      &  1.04463709 - 0.06546290i   & 1.04250023 - 0.19657035i \\ 
\hline
4      &   1.41041612 - 0.06615598i  &  1.40895830 - 0.19857604i \\ 
\hline
5       &   1.77187959 - 0.06663877i   & 1.77074409 - 0.19998893i  \\ 
\hline
\end{tabular}
\end{table}

\par We employ two publicly available codes---the \texttt{qnm} package \cite{Stein:2019mop} and the Julia package \texttt{KerrQuasinormalModes.jl} \cite{Asad:2024kqm}---to compute the QNM frequencies of the fundamental mode and the first overtone of gravitational perturbations ($s=-2$) for Kerr BHs over a wide range of parameters. The results obtained from the two codes agree with each other to within $10^{-10}$.\footnote{See \cite{Motohashi:2024fwt,Pombo:2025urp} for further methods for high-precision calculations of the QNMs of Kerr BHs.} The numerical values of the QNM frequencies are listed in Tables~\ref{qnm.3.6.9}–\ref{qnm2345} \cite{DataAvailabilityStatement}, where all values are reported to eight significant figures. Here, for simplicity we set $M=1$. 
\begin{figure*}
    \centering
    \begin{minipage}{0.48\textwidth}
        \centering
        \includegraphics[width=\linewidth]{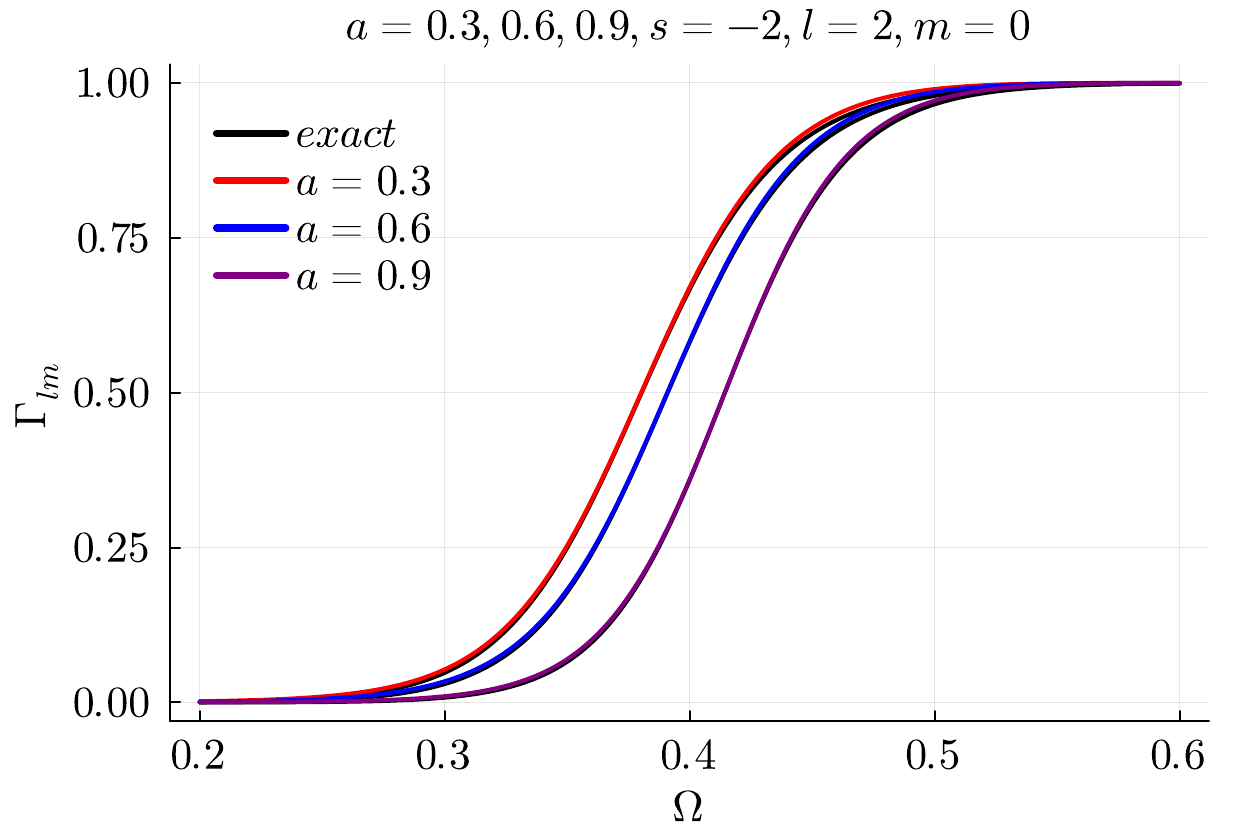}
    \end{minipage}
    \hfill  
    \begin{minipage}{0.48\textwidth}
        \centering
        \includegraphics[width=\linewidth]{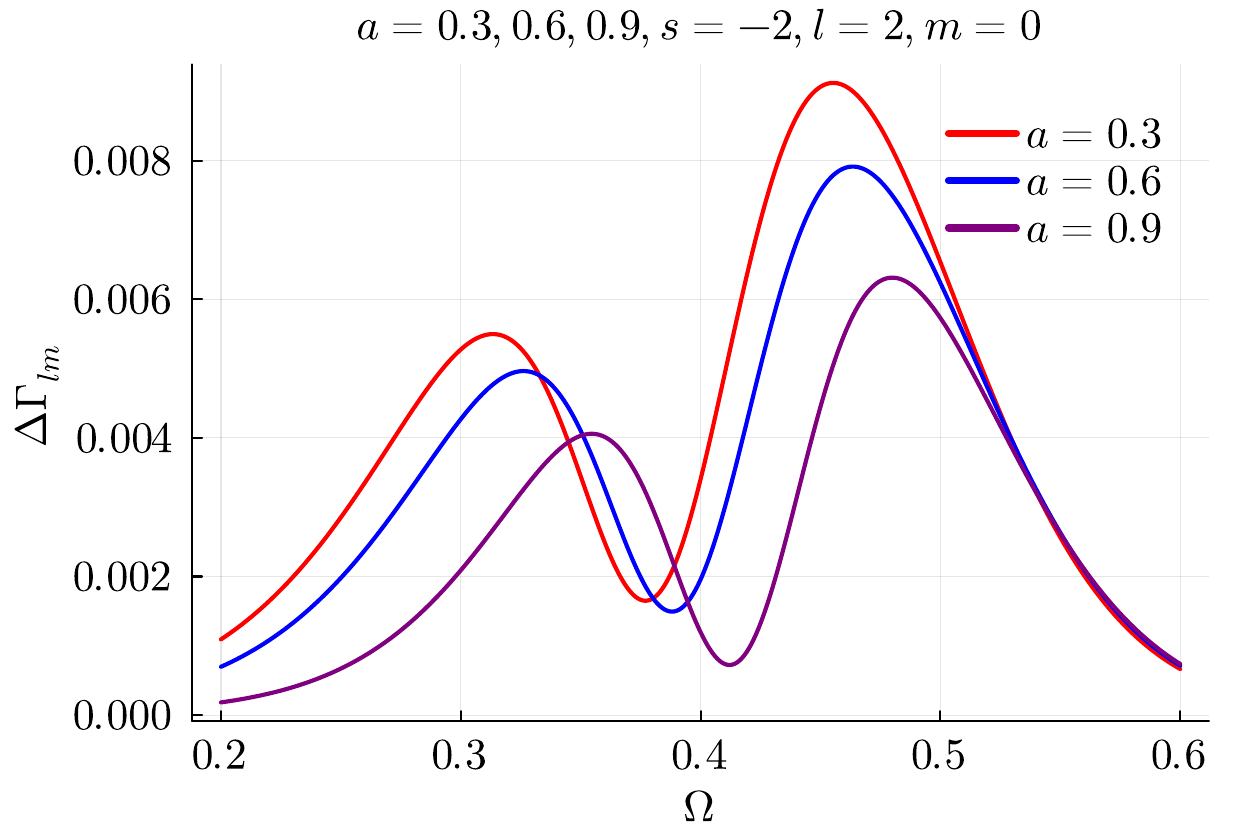}
    \end{minipage}
    \caption{Left: Exact GBFs (black lines) and the corresponding QNM/GBF-based approximations (colored lines) for  the gravitational perturbations of Kerr BHs with  spin parameters $a=0.3,0.6,0.9$, ordered from left to right. The results are shown for $l=2$ and $m=0$ as functions of the real frequency $\Omega$. 
    Right: The difference $\Delta\Gamma_{20}$ between the QNM/GBF-based approximations and the exact values  for the same  parameters.  }
\label{a.3.6.9l2m0}
\end{figure*}

\begin{figure*} 
    \centering
    \begin{minipage}{0.48\textwidth}
        \centering
        \includegraphics[width=\linewidth]{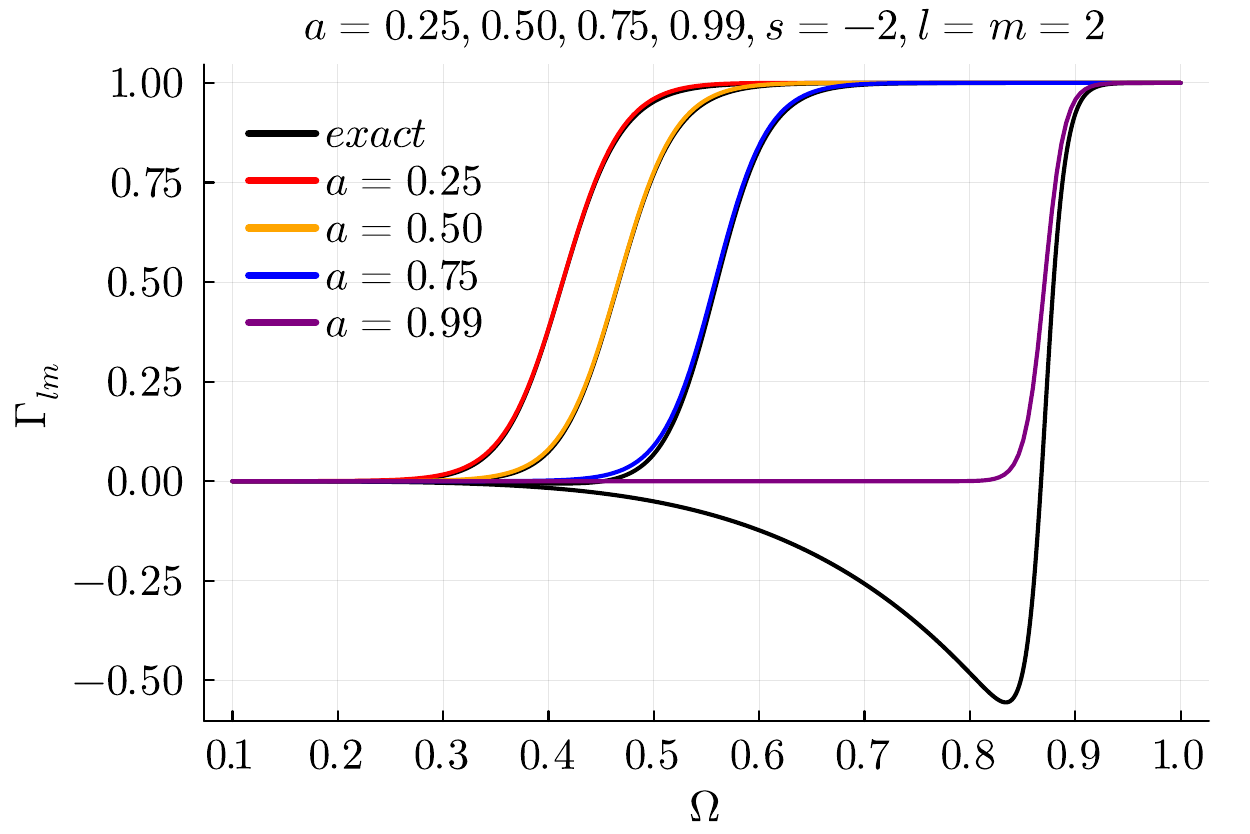}
    \end{minipage}
    \hfill  
    \begin{minipage}{0.48\textwidth}
        \centering
        \includegraphics[width=\linewidth]{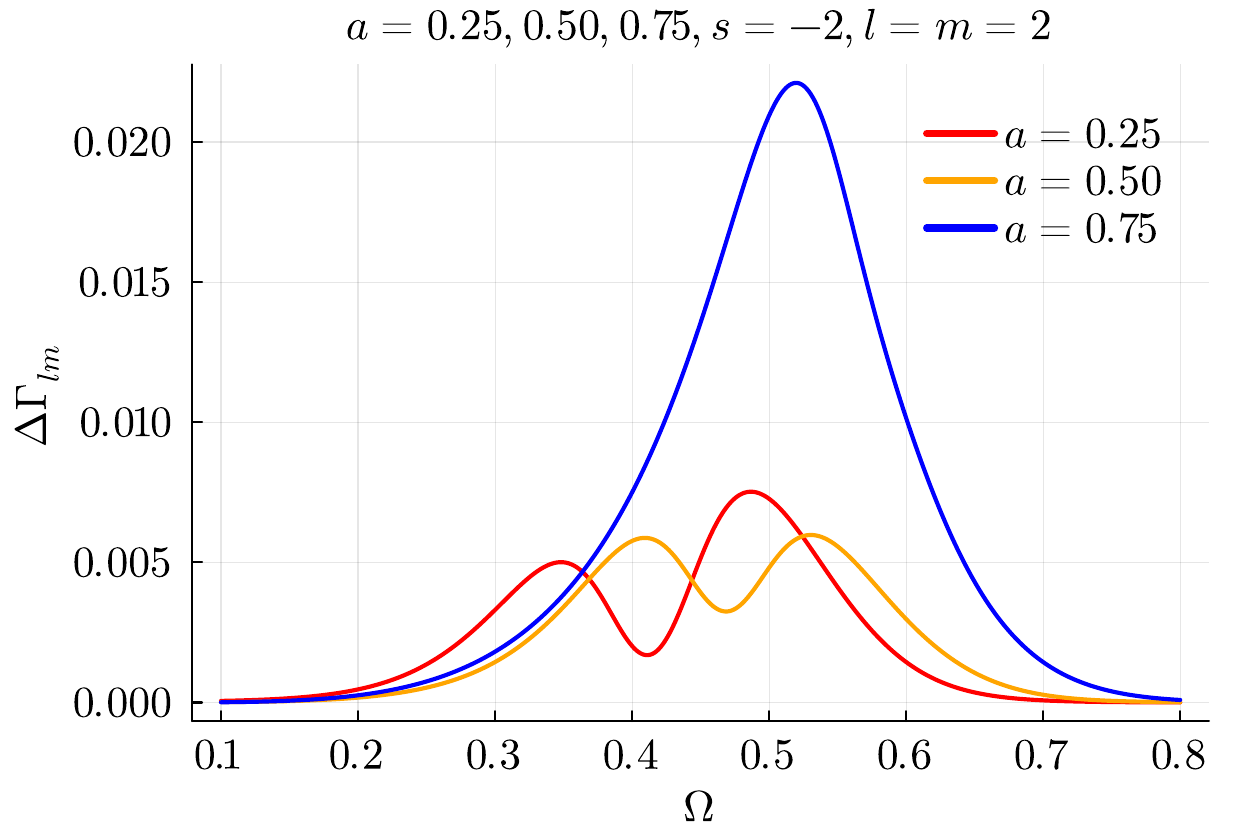}
    \end{minipage}
    \caption{Left: Same as in Fig. \ref{a.3.6.9l2m0}, but for GBFs with $l=m=2$, and  spin parameters $a=0.25,0.50,0.75,0.99$ ordered from left to right.  
   Right: The difference $\Delta\Gamma_{22}$ between the QNM/GBF-based approximations and the exact values for $a=0.25,0.50,0.75$.}  
\label{a.25.50.75.99lm2}
\end{figure*}

\begin{figure*}
    \centering
    \begin{minipage}{0.48\textwidth}
        \centering
        \includegraphics[width=\linewidth]{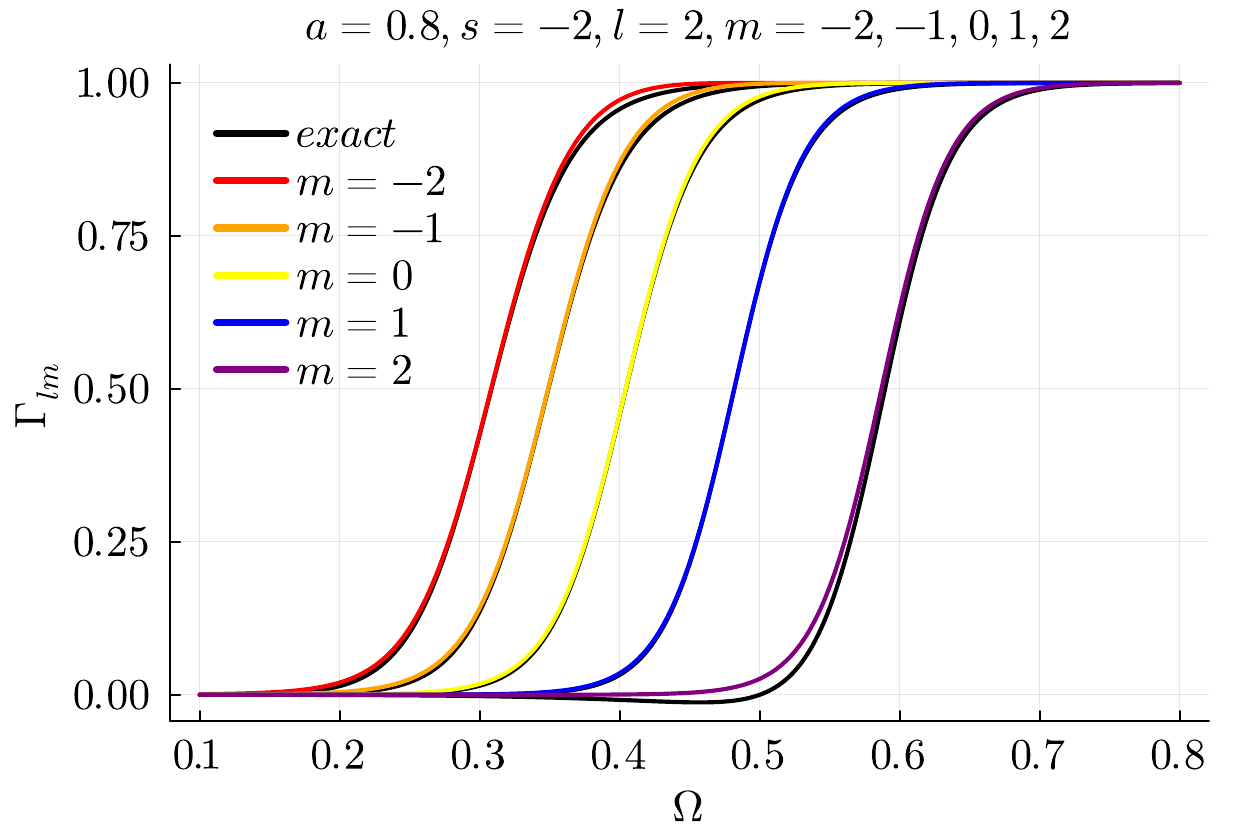}
    \end{minipage}
    \hfill  
    \begin{minipage}{0.48\textwidth}
        \centering
        \includegraphics[width=\linewidth]{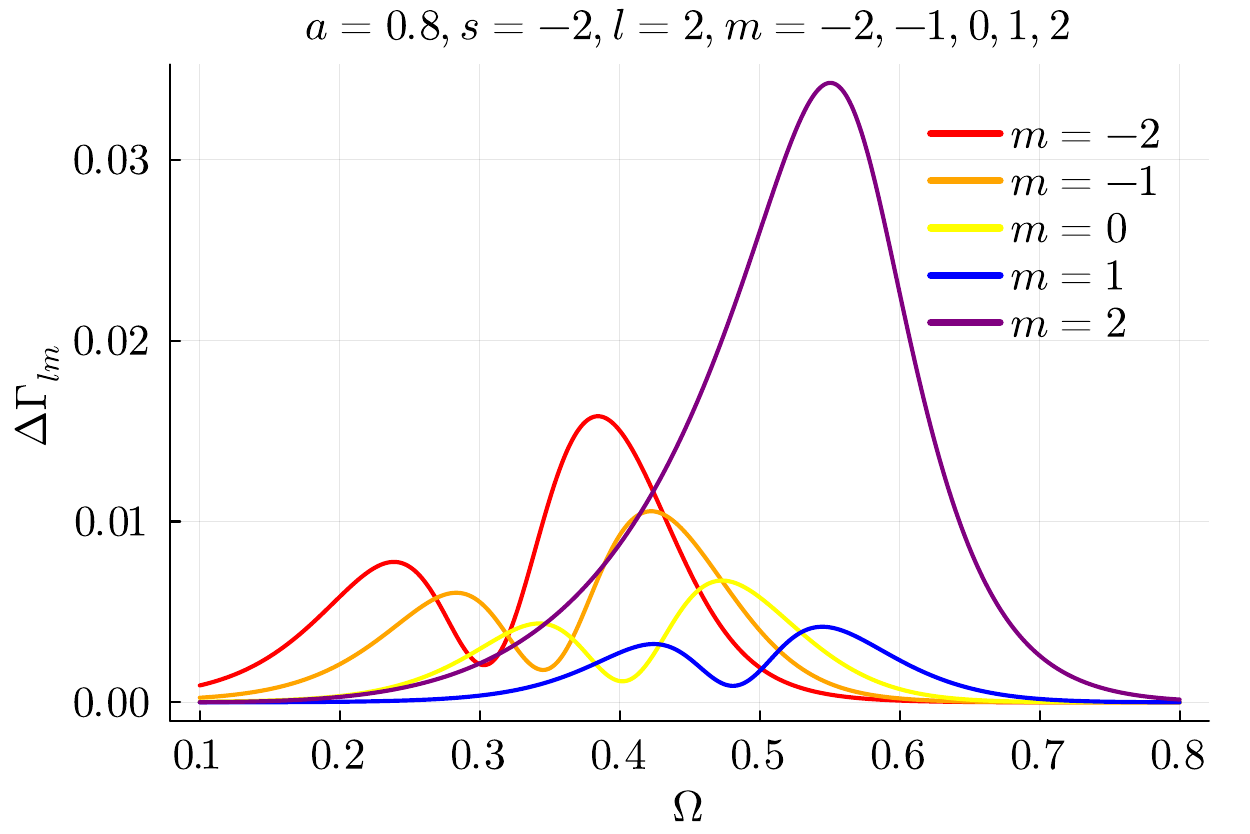}
    \end{minipage}
\caption{Left: Exact GBFs (black lines) and the corresponding QNM/GBF-based approximations  (colored lines) for the gravitational perturbations of Kerr BHs  with $a=0.8$, $l=2$ and $m=-2,-1,0,1,2$, ordered from left to right. Right: The difference $\Delta\Gamma_{2m}$ between QNM/GBF-based approximations and the exact values for the same  parameters.}
\label{a.8l2m-2-2}
\end{figure*}

\begin{figure}
    \centering
    \includegraphics[width=0.7\linewidth]{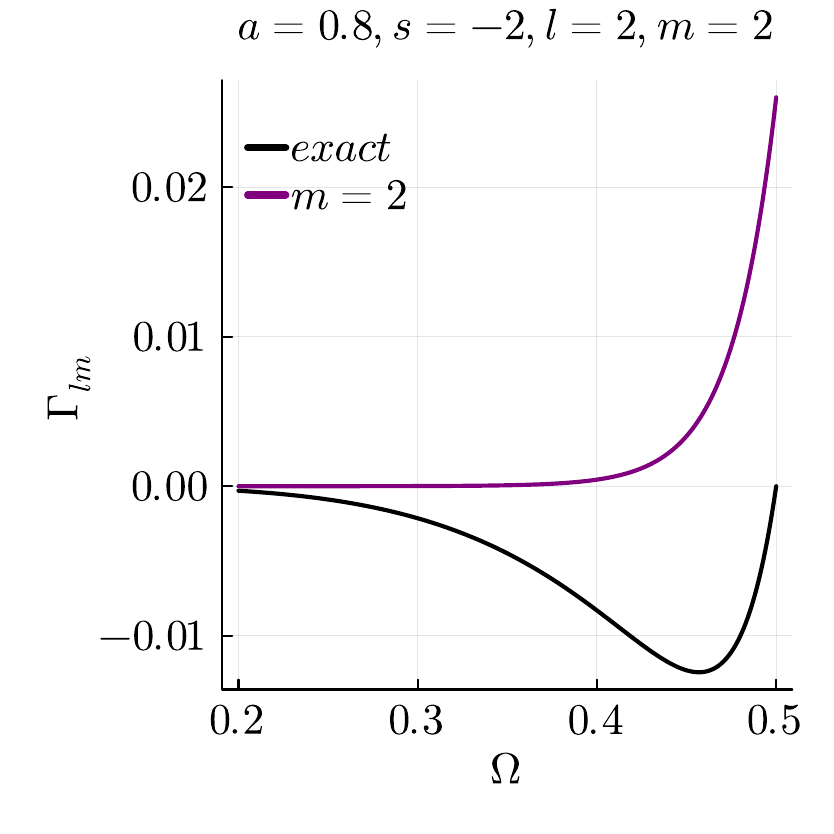}
    \caption{ This figure zooms into the superradiant frequency range of the $l=m=2$ mode from Fig. \ref{a.8l2m-2-2}, juxtaposing the exact GBFs (black line) with their approximations (purple line). The superradiant breakdown
of the correspondence becomes visible.}
    \label{a.8l2m2}
\end{figure}

\begin{figure*}
    \centering
    \begin{minipage}{0.48\textwidth}
        \centering
        \includegraphics[width=\linewidth]{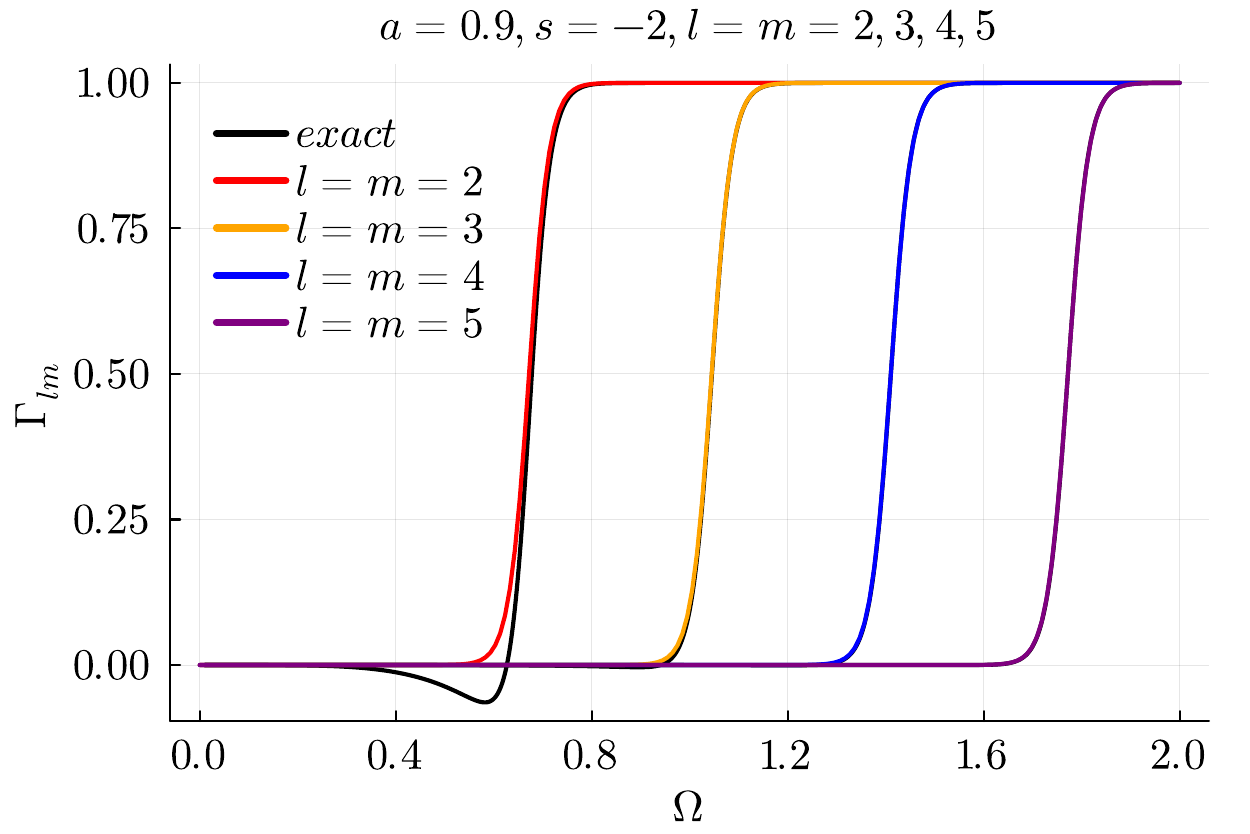}
    \end{minipage}
    \hfill  
    \begin{minipage}{0.48\textwidth}
        \centering
        \includegraphics[width=\linewidth]{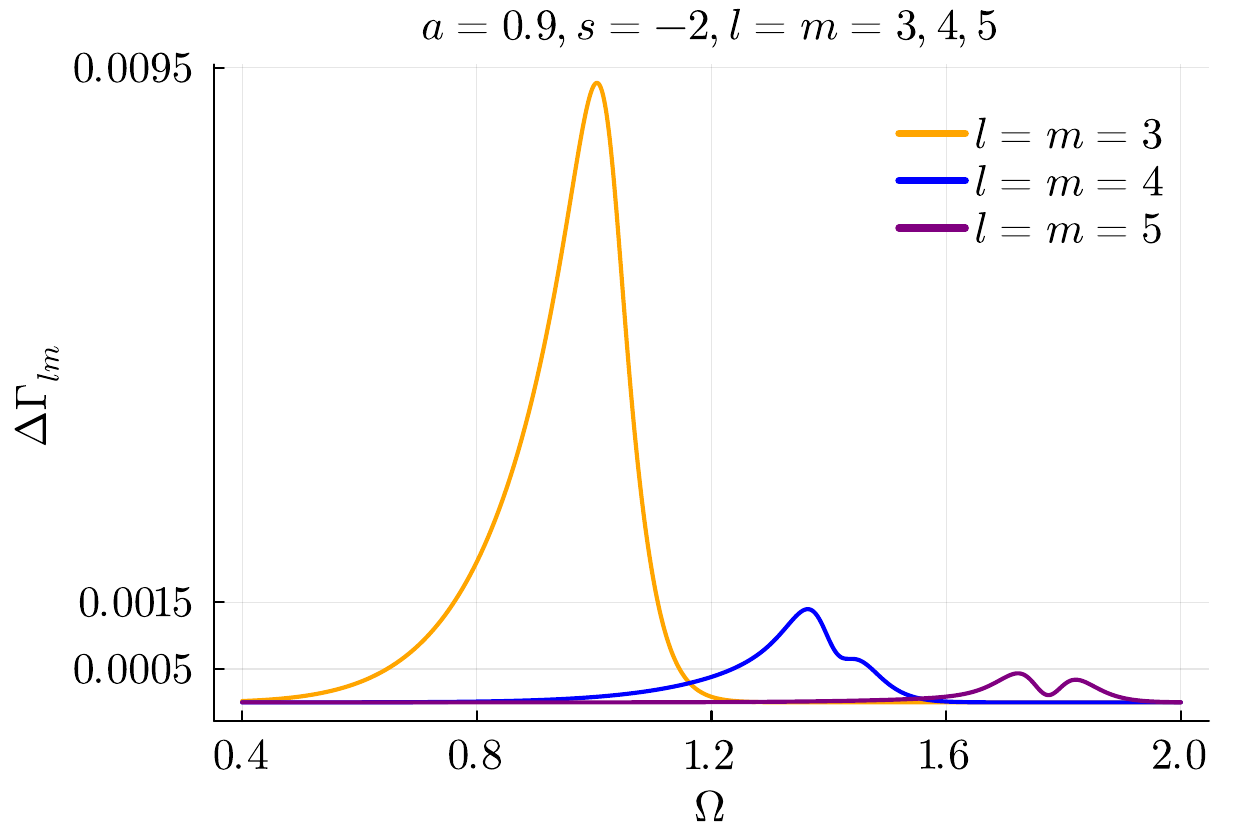}
    \end{minipage}
\caption{Left: Exact GBFs $\Gamma_{lm}$ (black lines) and the corresponding QNM/GBF-based approximations (colored lines) for  the gravitational perturbations of Kerr BHs with  $a=0.9$,  $l=m=2,3,4,5$ ordered from left to right. 
 Right: The difference $\Delta\Gamma_{lm}$ between QNM/GBF-based approximations and the exact values for $l=m=3,4,5$.}
\label{a.9lm2-5}
\end{figure*}

\subsection{Exact computation of GBFs}
\par To obtain the exact values of GBFs of gravitational perturbations, we numerically solve the GSN equation \eqref{GSNeqn} for a scattering process. The physical boundary conditions for the perturbation field $X$ are given by
\be
X = \left\{ \begin{array}{ll} 
Z_\text{i}^{\mathrm{SN}} e^{-i\omega r_*} + Z_\text{r}^{\mathrm{SN}} e^{i\omega r_*}, & r_* \to +\infty, \\
Z_\text{t}^{\mathrm{SN}}  e^{-ipr_*} , &r_* \to  -\inf.
\end{array} \right.
\label{snab}
\ee
The connection between  $Z_\text{i}^{\mathrm{SN}},Z_\text{r}^{\mathrm{SN}}  $ and  $Z_\text{i}^{\mathrm{T}},Z_\text{r}^{\mathrm{T}} $  are \cite{Sasaki:2003xr}
\begin{align}
Z_\text{i}^{\mathrm{T}}=-\frac{1}{4\o^2} Z_\text{i}^{\mathrm{SN}},\quad
Z_\text{r}^{\mathrm{T}}=-\frac{4\o^2}{c_0} Z_\text{r}^{\mathrm{SN}},
\end{align}
where
\be
c_0  =  -12i\omega M + \lambda( \lambda+2) -12 a\omega \left( a\omega - m\right).
\ee
Then the GBFs can be defined as the ratio of the total energy flux absorbed by the BH to the incident energy flux from infinity \cite{Teukolsky:1974yv,Brito:2015oca}
\begin{equation}
     \G_{lm} \equiv1- \left|\frac{C}{c_0}\right|^2 \left|\frac{Z_\text{r}^{\mathrm{SN}}}{Z_\text{i}^{\mathrm{SN}}} \right|^2 ,
\end{equation}
where the factor $|C|^2$ reads 

\begin{align}
    |C|^2=&B^2[(Q-2)^2+36ma\o -36a^2\o^2]\\
    &+48(2Q-1)(a^2\o^2-ma\o)+144\o^2(M^2-a^2),\nonumber
\end{align}
with
\be
 B^2=Q^2+4ma\o-4a^2\o^2,\quad  Q=\l+s(s+1).
\ee
\par First of all, the coupling constant $\lambda$ being as the eigenvalue of the angular Teukolsky equation (\ref{angularTeukolskyequation}) can be computed either through tools provided by \texttt{Black Hole Perturbations Toolkit} \cite{BHPToolkit} or the  public code \texttt{SpinWeightedSpheroidalHarmonics.jl}  \cite{Lo:2023fvv}. Then we compute the accurate GBFs by solving the GSN equation \eqref{GSNeqn} with the public code \texttt{GeneralizedSasakiNakamura.jl} \cite{Lo:2023fvv}
(with the same code, QNM frequencies can also be obtained using the Newton-Raphson method \cite{Lo:2025njp}).
The comparison of GBFs of the gravitational perturbations for Kerr BHs obtained from two different methods is shown in Figs. \ref{a.3.6.9l2m0}-\ref{a.9lm2-5}.

\par To quantify the accuracy of the QNM/GBF-based approximations relative to the exact GBFs, 
we define  the maximum absolute error over the plotted frequency range as
\be
\delta_{lm}\equiv\max_{\Omega}|\D\G_{lm}|=\max_{\Omega} |\G_{lm} - \G_{lm}^{\text{exact}}|,
\ee
where $\Gamma_{lm}$ denotes the QNM/GBF-based approximations, and $\G_{lm}^{\text{exact}}$ is the exact result. This quantity will be used below to assess the quality of the approximation for different values of $a$, $l$ and $m$.

\par We first  consider $(l,m)=(2,0)$ mode for which the superradiance does not occur. Fig. \ref{a.3.6.9l2m0} shows the comparison between exact GBFs and the approximations obtained via QNM/GBF correspondence for different values of the rotation parameter $a$. We observe that the correspondence provides a good approximation for the GBFs of the Kerr BH. The difference between the approximations and the exact GBF values is less than 0.01. More specifically, for the spin parameters $a=0.3,0.6,0.9$, the maximum absolute errors $\delta_{20}=0.0091,0.0079,0.0063$, respectively.

\par We also examine the GBFs for various values of the spin parameter $a$ for the $(l,m)=(2,2)$ mode, including the onset of the superradiant regime. As shown in Fig.~\ref{a.25.50.75.99lm2}, the correspondence still provides a good approximation for relatively slowly  rotating Kerr BHs.
The maximum absolute errors $\delta_{22}=$ $0.0075, 0.0060, 0.0221$ for $a=0.25,0.50,0.75$, respectively. Thus, except for the case $a=0.75$, the error remains below $0.01$,  similar to the behavior found for the $l=2$ and $m=0$ mode. For $a=0.75$, the frequency interval over which superradiance occurs becomes relatively wide, and the WKB-based formula correspondingly becomes less accurate. The superradiant effect is even more pronounced for the nearly extremal case $a=0.99$. In that regime, however, the correspondence formula breaks down, because by construction it yields only non-negative GBFs, whereas superradiance requires $\Gamma_{lm}<0$. Negative  GBFs in the superradiant regime is essential for accurately computing the Hawking spectrum of a rotating BH \cite{Pedrotti:2025idg}.

Figure~\ref{a.8l2m-2-2} presents the GBFs for rapidly rotating Kerr BHs with $a=0.8$, $l=2$ and $m=-2,-1,0,1,2$. The corresponding maximum absolute errors are $\delta_{2m}=0.0158, 0.0106, 0.0067, 0.0042, 0.0342$, for $m=-2,-1,0,1,2$, respectively. Overall, the QNM/GBF correspondence remains accurate in regimes where superradiance is absent or where the superradiant frequency interval is small. Moreover, the correspondence seems to perform better for modes with larger real parts of the QNM frequencies, which may account for the fact that $\delta_{21}<\delta_{2,-1}$ and the behavior in the previous two figures.  In Fig.~\ref{a.8l2m2}, we examine more closely the GBFs for the $l=m=2$ mode, where superradiance becomes pronounced and the correspondence breaks down. 

In Fig.~\ref{a.9lm2-5}, we show the GBFs for rapidly rotating Kerr BHs with $a=0.9$ and $l=m=2,3,4,5$. From this figure, we observe that for  $a=0.9$, and $l=m=3,4,5$, the maximum absolute errors $\delta_{lm}=0.0092$, $0.0014$, and $0.0004$, respectively. It is clear that $\delta_{lm}$ decreases as $l$ increases, indicating that the analytic formula becomes increasingly accurate at large $l$. This behavior is fully consistent with the WKB expectation that the approximation improves toward the eikonal limit, where it becomes exact.

A further remark concerns the near-extremal regime. As shown in Table \ref{qnm.25.50.75.99}, for $a=0.99$ and $l=m=2$, the fundamental mode and the first overtone become very close in their real parts. This, however, does not by itself invalidate the correction formulas in Eqs. \eqref{6thfm}-\eqref{delta3}.  The reason is that these formulas remain well defined even when $\omega_0$ and $\omega_1$ are close; their proximity does not introduce any formal singularity, as can be checked directly from Eqs. \eqref{6thfm}-\eqref{delta3}. In practice, the relevant issue is whether the two modes can still be extracted reliably. For the parameter range considered here, the QNM frequencies computed using two independent public codes agree to within $10^{-10}$, indicating that the identification of $\omega_0$ and $\omega_1$ remains numerically stable. The main limitation in the near-extremal/high-spin regime is instead associated with the approach to the superradiant regime, where the WKB-based GBF formula is no longer applicable.

\section{Conclusion}
\label{conclusion}
\par In this work, we have developed a systematic derivation of the correspondence between QNMs and GBFs for Kerr BHs within the WKB framework. By reformulating the radial Teukolsky equation into a Schr\"odinger-like equation with a short-ranged effective potential, we derived unified connection formulas that govern both the QNM spectrum and the scattering coefficients, and obtained explicit QNM/GBF correspondence relations including higher-order WKB corrections.

\par We tested the correspondence for gravitational perturbations by comparing the GBFs reconstructed from QNM frequencies with exact numerical results obtained from the generalized Sasaki–Nakamura equation. Excellent agreement is observed in nonsuperradiant regimes, with discrepancies decreasing rapidly as the angular quantum number increases, consistent with expectations from the eikonal approximation. The correspondence is found to break down in the superradiant regime, where the WKB-based formalism cannot capture the sign change of the GBFs.

Our results demonstrate that, within its domain of validity, the QNM/GBF correspondence provides a reliable and efficient bridge between dynamical and radiative properties of rotating BHs. The formalism developed here can be straightforwardly extended to other rotating spacetimes, including Kerr–Newman \cite{Li:2021zct} and higher-dimensional Kerr BHs \cite{Davey:2022vyx}. Possible future directions include computing GBFs from exact solutions \cite{Motohashi:2021zyv} of the Teukolsky equation and exploring alternative approaches based on Kerr/CFT correspondence \cite{Bredberg:2009pv}.

\section*{Acknowledgments}
\par We thank Xin-Dong Du for helpful discussions. This work makes use of the Black Hole Perturbations Toolkit. The work is in part supported by NSFC Grant No. 12205104 and the startup funding of South China University of Technology. 


\begin{thebibliography}{80}%
	\makeatletter
	\providecommand \@ifxundefined [1]{%
		\@ifx{#1\undefined}
	}%
	\providecommand \@ifnum [1]{%
		\ifnum #1\expandafter \@firstoftwo
		\else \expandafter \@secondoftwo
		\fi
	}%
	\providecommand \@ifx [1]{%
		\ifx #1\expandafter \@firstoftwo
		\else \expandafter \@secondoftwo
		\fi
	}%
	\providecommand \natexlab [1]{#1}%
	\providecommand \enquote  [1]{``#1''}%
	\providecommand \bibnamefont  [1]{#1}%
	\providecommand \bibfnamefont [1]{#1}%
	\providecommand \citenamefont [1]{#1}%
	\providecommand \href@noop [0]{\@secondoftwo}%
	\providecommand \href [0]{\begingroup \@sanitize@url \@href}%
	\providecommand \@href[1]{\@@startlink{#1}\@@href}%
	\providecommand \@@href[1]{\endgroup#1\@@endlink}%
	\providecommand \@sanitize@url [0]{\catcode `\\12\catcode `\$12\catcode
		`\&12\catcode `\#12\catcode `\^12\catcode `\_12\catcode `\%12\relax}%
	\providecommand \@@startlink[1]{}%
	\providecommand \@@endlink[0]{}%
	\providecommand \url  [0]{\begingroup\@sanitize@url \@url }%
	\providecommand \@url [1]{\endgroup\@href {#1}{\urlprefix }}%
	\providecommand \urlprefix  [0]{URL }%
	\providecommand \Eprint [0]{\href }%
	\providecommand \doibase [0]{http://dx.doi.org/}%
	\providecommand \selectlanguage [0]{\@gobble}%
	\providecommand \bibinfo  [0]{\@secondoftwo}%
	\providecommand \bibfield  [0]{\@secondoftwo}%
	\providecommand \translation [1]{[#1]}%
	\providecommand \BibitemOpen [0]{}%
	\providecommand \bibitemStop [0]{}%
	\providecommand \bibitemNoStop [0]{.\EOS\space}%
	\providecommand \EOS [0]{\spacefactor3000\relax}%
	\providecommand \BibitemShut  [1]{\csname bibitem#1\endcsname}%
	\let\auto@bib@innerbib\@empty
	\bibitem [{\citenamefont {Kokkotas}\ and\ \citenamefont
		{Schmidt}(1999)}]{Kokkotas:1999bd}%
	\BibitemOpen
	\bibfield  {author} {\bibinfo {author} {\bibfnamefont {K.~D.}\ \bibnamefont
			{Kokkotas}}\ and\ \bibinfo {author} {\bibfnamefont {B.~G.}\ \bibnamefont
			{Schmidt}},\ }\href {\doibase 10.12942/lrr-1999-2} {\bibfield  {journal}
		{\bibinfo  {journal} {Living Rev. Rel.}\ }\textbf {\bibinfo {volume} {2}},\
		\bibinfo {pages} {2} (\bibinfo {year} {1999})},\ \Eprint
	{http://arxiv.org/abs/gr-qc/9909058} {arXiv:gr-qc/9909058} \BibitemShut
	{NoStop}%
	\bibitem [{\citenamefont {Berti}\ \emph {et~al.}(2009)\citenamefont {Berti},
		\citenamefont {Cardoso},\ and\ \citenamefont {Starinets}}]{Berti:2009kk}%
	\BibitemOpen
	\bibfield  {author} {\bibinfo {author} {\bibfnamefont {E.}~\bibnamefont
			{Berti}}, \bibinfo {author} {\bibfnamefont {V.}~\bibnamefont {Cardoso}}, \
		and\ \bibinfo {author} {\bibfnamefont {A.~O.}\ \bibnamefont {Starinets}},\
	}\href {\doibase 10.1088/0264-9381/26/16/163001} {\bibfield  {journal}
		{\bibinfo  {journal} {Class. Quant. Grav.}\ }\textbf {\bibinfo {volume}
			{26}},\ \bibinfo {pages} {163001} (\bibinfo {year} {2009})},\ \Eprint
	{http://arxiv.org/abs/0905.2975} {arXiv:0905.2975 [gr-qc]} \BibitemShut
	{NoStop}%
	\bibitem [{\citenamefont {Konoplya}\ and\ \citenamefont
		{Zhidenko}(2011)}]{Konoplya:2011qq}%
	\BibitemOpen
	\bibfield  {author} {\bibinfo {author} {\bibfnamefont {R.~A.}\ \bibnamefont
			{Konoplya}}\ and\ \bibinfo {author} {\bibfnamefont {A.}~\bibnamefont
			{Zhidenko}},\ }\href {\doibase 10.1103/RevModPhys.83.793} {\bibfield
		{journal} {\bibinfo  {journal} {Rev. Mod. Phys.}\ }\textbf {\bibinfo {volume}
			{83}},\ \bibinfo {pages} {793} (\bibinfo {year} {2011})},\ \Eprint
	{http://arxiv.org/abs/1102.4014} {arXiv:1102.4014 [gr-qc]} \BibitemShut
	{NoStop}%
	\bibitem [{\citenamefont {Berti}\ \emph {et~al.}(2025)\citenamefont {Berti}
		\emph {et~al.}}]{Berti:2025hly}%
	\BibitemOpen
	\bibfield  {author} {\bibinfo {author} {\bibfnamefont {E.}~\bibnamefont
			{Berti}} \emph {et~al.},\ }\href@noop {} {\  (\bibinfo {year} {2025})},\
	\Eprint {http://arxiv.org/abs/2505.23895} {arXiv:2505.23895 [gr-qc]}
	\BibitemShut {NoStop}%
	\bibitem [{\citenamefont {Hawking}(1975)}]{Hawking:1975vcx}%
	\BibitemOpen
	\bibfield  {author} {\bibinfo {author} {\bibfnamefont {S.~W.}\ \bibnamefont
			{Hawking}},\ }\href {\doibase 10.1007/BF02345020} {\bibfield  {journal}
		{\bibinfo  {journal} {Commun. Math. Phys.}\ }\textbf {\bibinfo {volume}
			{43}},\ \bibinfo {pages} {199} (\bibinfo {year} {1975})},\ \bibinfo {note}
	{[Erratum: Commun.Math.Phys. 46, 206 (1976)]}\BibitemShut {NoStop}%
	\bibitem [{\citenamefont {Parikh}\ and\ \citenamefont
		{Wilczek}(2000)}]{PhysRevLett.85.5042}%
	\BibitemOpen
	\bibfield  {author} {\bibinfo {author} {\bibfnamefont {M.~K.}\ \bibnamefont
			{Parikh}}\ and\ \bibinfo {author} {\bibfnamefont {F.}~\bibnamefont
			{Wilczek}},\ }\href {\doibase 10.1103/PhysRevLett.85.5042} {\bibfield
		{journal} {\bibinfo  {journal} {Phys. Rev. Lett.}\ }\textbf {\bibinfo
			{volume} {85}},\ \bibinfo {pages} {5042} (\bibinfo {year}
		{2000})}\BibitemShut {NoStop}%
	\bibitem [{\citenamefont {Oshita}(2024)}]{Oshita:2023cjz}%
	\BibitemOpen
	\bibfield  {author} {\bibinfo {author} {\bibfnamefont {N.}~\bibnamefont
			{Oshita}},\ }\href {\doibase 10.1103/PhysRevD.109.104028} {\bibfield
		{journal} {\bibinfo  {journal} {Phys. Rev. D}\ }\textbf {\bibinfo {volume}
			{109}},\ \bibinfo {pages} {104028} (\bibinfo {year} {2024})},\ \Eprint
	{http://arxiv.org/abs/2309.05725} {arXiv:2309.05725 [gr-qc]} \BibitemShut
	{NoStop}%
	\bibitem [{\citenamefont {Okabayashi}\ and\ \citenamefont
		{Oshita}(2024)}]{Okabayashi:2024qbz}%
	\BibitemOpen
	\bibfield  {author} {\bibinfo {author} {\bibfnamefont {K.}~\bibnamefont
			{Okabayashi}}\ and\ \bibinfo {author} {\bibfnamefont {N.}~\bibnamefont
			{Oshita}},\ }\href {\doibase 10.1103/PhysRevD.110.064086} {\bibfield
		{journal} {\bibinfo  {journal} {Phys. Rev. D}\ }\textbf {\bibinfo {volume}
			{110}},\ \bibinfo {pages} {064086} (\bibinfo {year} {2024})},\ \Eprint
	{http://arxiv.org/abs/2403.17487} {arXiv:2403.17487 [gr-qc]} \BibitemShut
	{NoStop}%
	\bibitem [{\citenamefont {Rosato}\ \emph {et~al.}(2025)\citenamefont {Rosato},
		\citenamefont {Yi}, \citenamefont {Berti},\ and\ \citenamefont
		{Pani}}]{Rosato:2025ulx}%
	\BibitemOpen
	\bibfield  {author} {\bibinfo {author} {\bibfnamefont {R.~F.}\ \bibnamefont
			{Rosato}}, \bibinfo {author} {\bibfnamefont {S.}~\bibnamefont {Yi}}, \bibinfo
		{author} {\bibfnamefont {E.}~\bibnamefont {Berti}}, \ and\ \bibinfo {author}
		{\bibfnamefont {P.}~\bibnamefont {Pani}},\ }\href@noop {} {\  (\bibinfo
		{year} {2025})},\ \Eprint {http://arxiv.org/abs/2512.15877} {arXiv:2512.15877
		[gr-qc]} \BibitemShut {NoStop}%
	\bibitem [{\citenamefont {Nair}(2025)}]{Nair:2025anr}%
	\BibitemOpen
	\bibfield  {author} {\bibinfo {author} {\bibfnamefont {S.}~\bibnamefont
			{Nair}},\ }\href@noop {} {\  (\bibinfo {year} {2025})},\ \Eprint
	{http://arxiv.org/abs/2509.09986} {arXiv:2509.09986 [gr-qc]} \BibitemShut
	{NoStop}%
	\bibitem [{\citenamefont {Kyutoku}\ \emph {et~al.}(2023)\citenamefont
		{Kyutoku}, \citenamefont {Motohashi},\ and\ \citenamefont
		{Tanaka}}]{Kyutoku:2022gbr}%
	\BibitemOpen
	\bibfield  {author} {\bibinfo {author} {\bibfnamefont {K.}~\bibnamefont
			{Kyutoku}}, \bibinfo {author} {\bibfnamefont {H.}~\bibnamefont {Motohashi}},
		\ and\ \bibinfo {author} {\bibfnamefont {T.}~\bibnamefont {Tanaka}},\ }\href
	{\doibase 10.1103/PhysRevD.107.044012} {\bibfield  {journal} {\bibinfo
			{journal} {Phys. Rev. D}\ }\textbf {\bibinfo {volume} {107}},\ \bibinfo
		{pages} {044012} (\bibinfo {year} {2023})},\ \Eprint
	{http://arxiv.org/abs/2206.00671} {arXiv:2206.00671 [gr-qc]} \BibitemShut
	{NoStop}%
	\bibitem [{\citenamefont {Rosato}\ \emph {et~al.}(2024)\citenamefont {Rosato},
		\citenamefont {Destounis},\ and\ \citenamefont {Pani}}]{Rosato:2024arw}%
	\BibitemOpen
	\bibfield  {author} {\bibinfo {author} {\bibfnamefont {R.~F.}\ \bibnamefont
			{Rosato}}, \bibinfo {author} {\bibfnamefont {K.}~\bibnamefont {Destounis}}, \
		and\ \bibinfo {author} {\bibfnamefont {P.}~\bibnamefont {Pani}},\ }\href
	{\doibase 10.1103/PhysRevD.110.L121501} {\bibfield  {journal} {\bibinfo
			{journal} {Phys. Rev. D}\ }\textbf {\bibinfo {volume} {110}},\ \bibinfo
		{pages} {L121501} (\bibinfo {year} {2024})},\ \Eprint
	{http://arxiv.org/abs/2406.01692} {arXiv:2406.01692 [gr-qc]} \BibitemShut
	{NoStop}%
	\bibitem [{\citenamefont {Oshita}\ \emph {et~al.}(2024)\citenamefont {Oshita},
		\citenamefont {Takahashi},\ and\ \citenamefont {Mukohyama}}]{Oshita:2024fzf}%
	\BibitemOpen
	\bibfield  {author} {\bibinfo {author} {\bibfnamefont {N.}~\bibnamefont
			{Oshita}}, \bibinfo {author} {\bibfnamefont {K.}~\bibnamefont {Takahashi}}, \
		and\ \bibinfo {author} {\bibfnamefont {S.}~\bibnamefont {Mukohyama}},\ }\href
	{\doibase 10.1103/PhysRevD.110.084070} {\bibfield  {journal} {\bibinfo
			{journal} {Phys. Rev. D}\ }\textbf {\bibinfo {volume} {110}},\ \bibinfo
		{pages} {084070} (\bibinfo {year} {2024})},\ \Eprint
	{http://arxiv.org/abs/2406.04525} {arXiv:2406.04525 [gr-qc]} \BibitemShut
	{NoStop}%
	\bibitem [{\citenamefont {Xie}\ \emph {et~al.}(2025)\citenamefont {Xie},
		\citenamefont {Wu},\ and\ \citenamefont {Guo}}]{Xie:2025jbr}%
	\BibitemOpen
	\bibfield  {author} {\bibinfo {author} {\bibfnamefont {L.}~\bibnamefont
			{Xie}}, \bibinfo {author} {\bibfnamefont {L.-B.}\ \bibnamefont {Wu}}, \ and\
		\bibinfo {author} {\bibfnamefont {Z.-K.}\ \bibnamefont {Guo}},\ }\href
	{\doibase 10.1103/v3xt-r8nc} {\bibfield  {journal} {\bibinfo  {journal}
			{Phys. Rev. D}\ }\textbf {\bibinfo {volume} {112}},\ \bibinfo {pages}
		{024054} (\bibinfo {year} {2025})},\ \Eprint
	{http://arxiv.org/abs/2505.21303} {arXiv:2505.21303 [gr-qc]} \BibitemShut
	{NoStop}%
	\bibitem [{\citenamefont {Konoplya}\ and\ \citenamefont
		{Zhidenko}(2024)}]{Konoplya:2024lir}%
	\BibitemOpen
	\bibfield  {author} {\bibinfo {author} {\bibfnamefont {R.~A.}\ \bibnamefont
			{Konoplya}}\ and\ \bibinfo {author} {\bibfnamefont {A.}~\bibnamefont
			{Zhidenko}},\ }\href {\doibase 10.1088/1475-7516/2024/09/068} {\bibfield
		{journal} {\bibinfo  {journal} {JCAP}\ }\textbf {\bibinfo {volume} {09}},\
		\bibinfo {pages} {068} (\bibinfo {year} {2024})},\ \Eprint
	{http://arxiv.org/abs/2406.11694} {arXiv:2406.11694 [gr-qc]} \BibitemShut
	{NoStop}%
	\bibitem [{\citenamefont {Iyer}\ and\ \citenamefont
		{Will}(1987)}]{Iyer:1986np}%
	\BibitemOpen
	\bibfield  {author} {\bibinfo {author} {\bibfnamefont {S.}~\bibnamefont
			{Iyer}}\ and\ \bibinfo {author} {\bibfnamefont {C.~M.}\ \bibnamefont
			{Will}},\ }\href {\doibase 10.1103/PhysRevD.35.3621} {\bibfield  {journal}
		{\bibinfo  {journal} {Phys. Rev. D}\ }\textbf {\bibinfo {volume} {35}},\
		\bibinfo {pages} {3621} (\bibinfo {year} {1987})}\BibitemShut {NoStop}%
	\bibitem [{\citenamefont {Konoplya}\ \emph {et~al.}(2019)\citenamefont
		{Konoplya}, \citenamefont {Zhidenko},\ and\ \citenamefont
		{Zinhailo}}]{Konoplya:2019hlu}%
	\BibitemOpen
	\bibfield  {author} {\bibinfo {author} {\bibfnamefont {R.~A.}\ \bibnamefont
			{Konoplya}}, \bibinfo {author} {\bibfnamefont {A.}~\bibnamefont {Zhidenko}},
		\ and\ \bibinfo {author} {\bibfnamefont {A.~F.}\ \bibnamefont {Zinhailo}},\
	}\href {\doibase 10.1088/1361-6382/ab2e25} {\bibfield  {journal} {\bibinfo
			{journal} {Class. Quant. Grav.}\ }\textbf {\bibinfo {volume} {36}},\ \bibinfo
		{pages} {155002} (\bibinfo {year} {2019})},\ \Eprint
	{http://arxiv.org/abs/1904.10333} {arXiv:1904.10333 [gr-qc]} \BibitemShut
	{NoStop}%
	\bibitem [{\citenamefont {Konoplya}\ and\ \citenamefont
		{Zhidenko}(2025)}]{Konoplya:2024vuj}%
	\BibitemOpen
	\bibfield  {author} {\bibinfo {author} {\bibfnamefont {R.~A.}\ \bibnamefont
			{Konoplya}}\ and\ \bibinfo {author} {\bibfnamefont {A.}~\bibnamefont
			{Zhidenko}},\ }\href {\doibase 10.1016/j.physletb.2025.139288} {\bibfield
		{journal} {\bibinfo  {journal} {Phys. Lett. B}\ }\textbf {\bibinfo {volume}
			{861}},\ \bibinfo {pages} {139288} (\bibinfo {year} {2025})},\ \Eprint
	{http://arxiv.org/abs/2408.11162} {arXiv:2408.11162 [gr-qc]} \BibitemShut
	{NoStop}%
	\bibitem [{\citenamefont {Pedrotti}\ and\ \citenamefont
		{Calz{\`a}}(2025)}]{Pedrotti:2025idg}%
	\BibitemOpen
	\bibfield  {author} {\bibinfo {author} {\bibfnamefont {D.}~\bibnamefont
			{Pedrotti}}\ and\ \bibinfo {author} {\bibfnamefont {M.}~\bibnamefont
			{Calz{\`a}}},\ }\href {\doibase 10.1103/1q35-mjjz} {\bibfield  {journal}
		{\bibinfo  {journal} {Phys. Rev. D}\ }\textbf {\bibinfo {volume} {111}},\
		\bibinfo {pages} {124056} (\bibinfo {year} {2025})},\ \Eprint
	{http://arxiv.org/abs/2504.01909} {arXiv:2504.01909 [gr-qc]} \BibitemShut
	{NoStop}%
	\bibitem [{\citenamefont {Malik}(2025{\natexlab{a}})}]{Malik:2024cgb}%
	\BibitemOpen
	\bibfield  {author} {\bibinfo {author} {\bibfnamefont {Z.}~\bibnamefont
			{Malik}},\ }\href {\doibase 10.1088/1475-7516/2025/04/042} {\bibfield
		{journal} {\bibinfo  {journal} {JCAP}\ }\textbf {\bibinfo {volume} {04}},\
		\bibinfo {pages} {042} (\bibinfo {year} {2025}{\natexlab{a}})},\ \Eprint
	{http://arxiv.org/abs/2412.19443} {arXiv:2412.19443 [gr-qc]} \BibitemShut
	{NoStop}%
	\bibitem [{\citenamefont {Malik}(2025{\natexlab{b}})}]{Malik:2025erb}%
	\BibitemOpen
	\bibfield  {author} {\bibinfo {author} {\bibfnamefont {Z.}~\bibnamefont
			{Malik}},\ }\href {\doibase 10.53941/ijgtp.2025.100006} {\bibfield  {journal}
		{\bibinfo  {journal} {Theor. Phys.}\ }\textbf {\bibinfo {volume} {1}},\
		\bibinfo {pages} {6} (\bibinfo {year} {2025}{\natexlab{b}})},\ \Eprint
	{http://arxiv.org/abs/2509.15995} {arXiv:2509.15995 [gr-qc]} \BibitemShut
	{NoStop}%
	\bibitem [{\citenamefont {Heidari}\ \emph {et~al.}(2024)\citenamefont
		{Heidari}, \citenamefont {Ara{\'u}jo~Filho}, \citenamefont {Vertogradov},\
		and\ \citenamefont {{\"O}vg{\"u}n}}]{Heidari:2024bbd}%
	\BibitemOpen
	\bibfield  {author} {\bibinfo {author} {\bibfnamefont {N.}~\bibnamefont
			{Heidari}}, \bibinfo {author} {\bibfnamefont {A.~A.}\ \bibnamefont
			{Ara{\'u}jo~Filho}}, \bibinfo {author} {\bibfnamefont {V.}~\bibnamefont
			{Vertogradov}}, \ and\ \bibinfo {author} {\bibfnamefont {A.}~\bibnamefont
			{{\"O}vg{\"u}n}},\ }\href@noop {} {\  (\bibinfo {year} {2024})},\ \Eprint
	{http://arxiv.org/abs/2412.05072} {arXiv:2412.05072 [gr-qc]} \BibitemShut
	{NoStop}%
	\bibitem [{\citenamefont {Skvortsova}(2024)}]{Skvortsova:2024msa}%
	\BibitemOpen
	\bibfield  {author} {\bibinfo {author} {\bibfnamefont {M.}~\bibnamefont
			{Skvortsova}},\ }\href@noop {} {\  (\bibinfo {year} {2024})},\ \Eprint
	{http://arxiv.org/abs/2411.06007} {arXiv:2411.06007 [gr-qc]} \BibitemShut
	{NoStop}%
	\bibitem [{\citenamefont {Dubinsky}(2025{\natexlab{a}})}]{Dubinsky:2025wns}%
	\BibitemOpen
	\bibfield  {author} {\bibinfo {author} {\bibfnamefont {A.}~\bibnamefont
			{Dubinsky}},\ }\href@noop {} {\  (\bibinfo {year} {2025}{\natexlab{a}})},\
	\Eprint {http://arxiv.org/abs/2511.00778} {arXiv:2511.00778 [gr-qc]}
	\BibitemShut {NoStop}%
	\bibitem [{\citenamefont {Tang}\ \emph
		{et~al.}(2025{\natexlab{a}})\citenamefont {Tang}, \citenamefont {Ling},\ and\
		\citenamefont {Jiang}}]{Tang:2025mkk}%
	\BibitemOpen
	\bibfield  {author} {\bibinfo {author} {\bibfnamefont {C.}~\bibnamefont
			{Tang}}, \bibinfo {author} {\bibfnamefont {Y.}~\bibnamefont {Ling}}, \ and\
		\bibinfo {author} {\bibfnamefont {Q.-Q.}\ \bibnamefont {Jiang}},\ }\href@noop
	{} {\  (\bibinfo {year} {2025}{\natexlab{a}})},\ \Eprint
	{http://arxiv.org/abs/2503.21597} {arXiv:2503.21597 [gr-qc]} \BibitemShut
	{NoStop}%
	\bibitem [{\citenamefont {L\"utf\"uo\u{g}lu}(2025)}]{Lutfuoglu:2025ohb}%
	\BibitemOpen
	\bibfield  {author} {\bibinfo {author} {\bibfnamefont {B.~C.}\ \bibnamefont
			{L\"utf\"uo\u{g}lu}},\ }\href@noop {} {\  (\bibinfo {year} {2025})},\ \Eprint
	{http://arxiv.org/abs/2505.06966} {arXiv:2505.06966 [gr-qc]} \BibitemShut
	{NoStop}%
	\bibitem [{\citenamefont {Shi}\ \emph {et~al.}(2025)\citenamefont {Shi},
		\citenamefont {Wang}, \citenamefont {Xiong},\ and\ \citenamefont
		{Li}}]{Shi:2025gst}%
	\BibitemOpen
	\bibfield  {author} {\bibinfo {author} {\bibfnamefont {Q.-L.}\ \bibnamefont
			{Shi}}, \bibinfo {author} {\bibfnamefont {R.}~\bibnamefont {Wang}}, \bibinfo
		{author} {\bibfnamefont {W.}~\bibnamefont {Xiong}}, \ and\ \bibinfo {author}
		{\bibfnamefont {P.-C.}\ \bibnamefont {Li}},\ }\href@noop {} {\  (\bibinfo
		{year} {2025})},\ \Eprint {http://arxiv.org/abs/2506.16217} {arXiv:2506.16217
		[gr-qc]} \BibitemShut {NoStop}%
	\bibitem [{\citenamefont {Bolokhov}\ and\ \citenamefont
		{Skvortsova}(2025{\natexlab{a}})}]{Bolokhov:2025lnt}%
	\BibitemOpen
	\bibfield  {author} {\bibinfo {author} {\bibfnamefont {S.~V.}\ \bibnamefont
			{Bolokhov}}\ and\ \bibinfo {author} {\bibfnamefont {M.}~\bibnamefont
			{Skvortsova}},\ }\href@noop {} {\bibfield  {journal} {\bibinfo  {journal}
			{Int. J. Grav. Theor. Phys.}\ }\textbf {\bibinfo {volume} {1}},\ \bibinfo
		{pages} {3} (\bibinfo {year} {2025}{\natexlab{a}})},\ \Eprint
	{http://arxiv.org/abs/2507.07196} {arXiv:2507.07196 [gr-qc]} \BibitemShut
	{NoStop}%
	\bibitem [{\citenamefont {Malik}(2025{\natexlab{c}})}]{Malik:2025dxn}%
	\BibitemOpen
	\bibfield  {author} {\bibinfo {author} {\bibfnamefont {Z.}~\bibnamefont
			{Malik}},\ }\href@noop {} {\  (\bibinfo {year} {2025}{\natexlab{c}})},\
	\Eprint {http://arxiv.org/abs/2508.19178} {arXiv:2508.19178 [gr-qc]}
	\BibitemShut {NoStop}%
	\bibitem [{\citenamefont {Dubinsky}(2025{\natexlab{b}})}]{Dubinsky:2025nxv}%
	\BibitemOpen
	\bibfield  {author} {\bibinfo {author} {\bibfnamefont {A.}~\bibnamefont
			{Dubinsky}},\ }\href@noop {} {\  (\bibinfo {year} {2025}{\natexlab{b}})},\
	\Eprint {http://arxiv.org/abs/2509.11017} {arXiv:2509.11017 [gr-qc]}
	\BibitemShut {NoStop}%
	\bibitem [{\citenamefont {L{\"u}tf{\"u}o{\u{g}}lu}\ \emph
		{et~al.}(2025)\citenamefont {L{\"u}tf{\"u}o{\u{g}}lu}, \citenamefont {Saka},
		\citenamefont {Shermatov}, \citenamefont {Rayimbaev}, \citenamefont
		{Ibragimov},\ and\ \citenamefont {Muminov}}]{Lutfuoglu:2025blw}%
	\BibitemOpen
	\bibfield  {author} {\bibinfo {author} {\bibfnamefont {B.~C.}\ \bibnamefont
			{L{\"u}tf{\"u}o{\u{g}}lu}}, \bibinfo {author} {\bibfnamefont {E.~U.}\
			\bibnamefont {Saka}}, \bibinfo {author} {\bibfnamefont {A.}~\bibnamefont
			{Shermatov}}, \bibinfo {author} {\bibfnamefont {J.}~\bibnamefont
			{Rayimbaev}}, \bibinfo {author} {\bibfnamefont {I.}~\bibnamefont
			{Ibragimov}}, \ and\ \bibinfo {author} {\bibfnamefont {S.}~\bibnamefont
			{Muminov}},\ }\href@noop {} {\  (\bibinfo {year} {2025})},\ \Eprint
	{http://arxiv.org/abs/2509.15923} {arXiv:2509.15923 [gr-qc]} \BibitemShut
	{NoStop}%
	\bibitem [{\citenamefont {Malik}(2025{\natexlab{d}})}]{Malik:2025qnr}%
	\BibitemOpen
	\bibfield  {author} {\bibinfo {author} {\bibfnamefont {Z.}~\bibnamefont
			{Malik}},\ }\href@noop {} {\  (\bibinfo {year} {2025}{\natexlab{d}})},\
	\Eprint {http://arxiv.org/abs/2510.06689} {arXiv:2510.06689 [gr-qc]}
	\BibitemShut {NoStop}%
	\bibitem [{\citenamefont {Dubinsky}(2025{\natexlab{c}})}]{Dubinsky:2025ypj}%
	\BibitemOpen
	\bibfield  {author} {\bibinfo {author} {\bibfnamefont {A.}~\bibnamefont
			{Dubinsky}},\ }\href@noop {} {\  (\bibinfo {year} {2025}{\natexlab{c}})},\
	\Eprint {http://arxiv.org/abs/2510.11643} {arXiv:2510.11643 [gr-qc]}
	\BibitemShut {NoStop}%
	\bibitem [{\citenamefont {Han}\ and\ \citenamefont {Gwak}(2025)}]{Han:2025cal}%
	\BibitemOpen
	\bibfield  {author} {\bibinfo {author} {\bibfnamefont {H.}~\bibnamefont
			{Han}}\ and\ \bibinfo {author} {\bibfnamefont {B.}~\bibnamefont {Gwak}},\
	}\href@noop {} {\  (\bibinfo {year} {2025})},\ \Eprint
	{http://arxiv.org/abs/2508.12989} {arXiv:2508.12989 [gr-qc]} \BibitemShut
	{NoStop}%
	\bibitem [{\citenamefont {Dubinsky}(2025{\natexlab{d}})}]{Dubinsky:2024vbn}%
	\BibitemOpen
	\bibfield  {author} {\bibinfo {author} {\bibfnamefont {A.}~\bibnamefont
			{Dubinsky}},\ }\href {\doibase 10.1142/S0217732325501111} {\bibfield
		{journal} {\bibinfo  {journal} {Mod. Phys. Lett. A}\ }\textbf {\bibinfo
			{volume} {40}},\ \bibinfo {pages} {2550111} (\bibinfo {year}
		{2025}{\natexlab{d}})},\ \Eprint {http://arxiv.org/abs/2412.00625}
	{arXiv:2412.00625 [gr-qc]} \BibitemShut {NoStop}%
	\bibitem [{\citenamefont {Ara{\'u}jo~Filho}(2025)}]{AraujoFilho:2025hkm}%
	\BibitemOpen
	\bibfield  {author} {\bibinfo {author} {\bibfnamefont {A.~A.}\ \bibnamefont
			{Ara{\'u}jo~Filho}},\ }\href {\doibase 10.1088/1475-7516/2025/06/026}
	{\bibfield  {journal} {\bibinfo  {journal} {JCAP}\ }\textbf {\bibinfo
			{volume} {06}},\ \bibinfo {pages} {026} (\bibinfo {year} {2025})},\ \Eprint
	{http://arxiv.org/abs/2501.00927} {arXiv:2501.00927 [gr-qc]} \BibitemShut
	{NoStop}%
	\bibitem [{\citenamefont {Heidari}\ and\ \citenamefont
		{Ara{\'u}jo~Filho}(2025)}]{Heidari:2025oop}%
	\BibitemOpen
	\bibfield  {author} {\bibinfo {author} {\bibfnamefont {N.}~\bibnamefont
			{Heidari}}\ and\ \bibinfo {author} {\bibfnamefont {A.~A.}\ \bibnamefont
			{Ara{\'u}jo~Filho}},\ }\href@noop {} {\  (\bibinfo {year} {2025})},\ \Eprint
	{http://arxiv.org/abs/2512.08604} {arXiv:2512.08604 [gr-qc]} \BibitemShut
	{NoStop}%
	\bibitem [{\citenamefont {L{\"u}tf{\"u}o{\u{g}}lu}(2025)}]{Lutfuoglu:2025ldc}%
	\BibitemOpen
	\bibfield  {author} {\bibinfo {author} {\bibfnamefont {B.~C.}\ \bibnamefont
			{L{\"u}tf{\"u}o{\u{g}}lu}},\ }\href {\doibase 10.53941/ijgtp.2025.100004}
	{\bibfield  {journal} {\bibinfo  {journal} {Theor. Phys.}\ }\textbf {\bibinfo
			{volume} {1}},\ \bibinfo {pages} {4} (\bibinfo {year} {2025})},\ \Eprint
	{http://arxiv.org/abs/2507.09246} {arXiv:2507.09246 [gr-qc]} \BibitemShut
	{NoStop}%
	\bibitem [{\citenamefont {Hamil}\ \emph {et~al.}(2025)\citenamefont {Hamil},
		\citenamefont {Al-Badawi},\ and\ \citenamefont
		{L{\"u}tf{\"u}o{\u{g}}lu}}]{Hamil:2025pte}%
	\BibitemOpen
	\bibfield  {author} {\bibinfo {author} {\bibfnamefont {B.}~\bibnamefont
			{Hamil}}, \bibinfo {author} {\bibfnamefont {A.}~\bibnamefont {Al-Badawi}}, \
		and\ \bibinfo {author} {\bibfnamefont {B.~C.}\ \bibnamefont
			{L{\"u}tf{\"u}o{\u{g}}lu}},\ }\href@noop {} {\  (\bibinfo {year} {2025})},\
	\Eprint {http://arxiv.org/abs/2505.18611} {arXiv:2505.18611 [gr-qc]}
	\BibitemShut {NoStop}%
	\bibitem [{\citenamefont {Yan}\ \emph {et~al.}(2025)\citenamefont {Yan},
		\citenamefont {Zhang}, \citenamefont {Yue},\ and\ \citenamefont
		{Li}}]{Yan:2025pvp}%
	\BibitemOpen
	\bibfield  {author} {\bibinfo {author} {\bibfnamefont {H.-P.}\ \bibnamefont
			{Yan}}, \bibinfo {author} {\bibfnamefont {Z.-Y.}\ \bibnamefont {Zhang}},
		\bibinfo {author} {\bibfnamefont {X.-J.}\ \bibnamefont {Yue}}, \ and\
		\bibinfo {author} {\bibfnamefont {X.-Q.}\ \bibnamefont {Li}},\ }\href@noop {}
	{\  (\bibinfo {year} {2025})},\ \Eprint {http://arxiv.org/abs/2511.21205}
	{arXiv:2511.21205 [gr-qc]} \BibitemShut {NoStop}%
	\bibitem [{\citenamefont {Konoplya}\ \emph {et~al.}(2025)\citenamefont
		{Konoplya}, \citenamefont {Khrabustovskyi}, \citenamefont
		{K{\v{r}}{\'\i}{\v{z}}},\ and\ \citenamefont {Zhidenko}}]{Konoplya:2025mvj}%
	\BibitemOpen
	\bibfield  {author} {\bibinfo {author} {\bibfnamefont {R.~A.}\ \bibnamefont
			{Konoplya}}, \bibinfo {author} {\bibfnamefont {A.}~\bibnamefont
			{Khrabustovskyi}}, \bibinfo {author} {\bibfnamefont {J.}~\bibnamefont
			{K{\v{r}}{\'\i}{\v{z}}}}, \ and\ \bibinfo {author} {\bibfnamefont
			{A.}~\bibnamefont {Zhidenko}},\ }\href {\doibase
		10.1088/1475-7516/2025/04/062} {\bibfield  {journal} {\bibinfo  {journal}
			{JCAP}\ }\textbf {\bibinfo {volume} {04}},\ \bibinfo {pages} {062} (\bibinfo
		{year} {2025})},\ \Eprint {http://arxiv.org/abs/2501.16134} {arXiv:2501.16134
		[gr-qc]} \BibitemShut {NoStop}%
	\bibitem [{\citenamefont {Sajadi}\ \emph {et~al.}(2025)\citenamefont {Sajadi},
		\citenamefont {Ponglertsakul},\ and\ \citenamefont {Gogoi}}]{Sajadi:2025kah}%
	\BibitemOpen
	\bibfield  {author} {\bibinfo {author} {\bibfnamefont {S.~N.}\ \bibnamefont
			{Sajadi}}, \bibinfo {author} {\bibfnamefont {S.}~\bibnamefont
			{Ponglertsakul}}, \ and\ \bibinfo {author} {\bibfnamefont {D.~J.}\
			\bibnamefont {Gogoi}},\ }\href {\doibase 10.1140/epjc/s10052-025-14555-6}
	{\bibfield  {journal} {\bibinfo  {journal} {Eur. Phys. J. C}\ }\textbf
		{\bibinfo {volume} {85}},\ \bibinfo {pages} {943} (\bibinfo {year} {2025})},\
	\Eprint {http://arxiv.org/abs/2503.18289} {arXiv:2503.18289 [gr-qc]}
	\BibitemShut {NoStop}%
	\bibitem [{\citenamefont {Malik}(2024)}]{Malik:2024wvs}%
	\BibitemOpen
	\bibfield  {author} {\bibinfo {author} {\bibfnamefont {Z.}~\bibnamefont
			{Malik}},\ }\href@noop {} {\  (\bibinfo {year} {2024})},\ \Eprint
	{http://arxiv.org/abs/2412.13385} {arXiv:2412.13385 [gr-qc]} \BibitemShut
	{NoStop}%
	\bibitem [{\citenamefont {Bolokhov}\ and\ \citenamefont
		{Skvortsova}(2025{\natexlab{b}})}]{Bolokhov:2024otn}%
	\BibitemOpen
	\bibfield  {author} {\bibinfo {author} {\bibfnamefont {S.~V.}\ \bibnamefont
			{Bolokhov}}\ and\ \bibinfo {author} {\bibfnamefont {M.}~\bibnamefont
			{Skvortsova}},\ }\href {\doibase 10.1088/1475-7516/2025/04/025} {\bibfield
		{journal} {\bibinfo  {journal} {JCAP}\ }\textbf {\bibinfo {volume} {04}},\
		\bibinfo {pages} {025} (\bibinfo {year} {2025}{\natexlab{b}})},\ \Eprint
	{http://arxiv.org/abs/2412.11166} {arXiv:2412.11166 [gr-qc]} \BibitemShut
	{NoStop}%
	\bibitem [{\citenamefont {Teukolsky}(1972)}]{Teukolsky:1972my}%
	\BibitemOpen
	\bibfield  {author} {\bibinfo {author} {\bibfnamefont {S.~A.}\ \bibnamefont
			{Teukolsky}},\ }\href {\doibase 10.1103/PhysRevLett.29.1114} {\bibfield
		{journal} {\bibinfo  {journal} {Phys. Rev. Lett.}\ }\textbf {\bibinfo
			{volume} {29}},\ \bibinfo {pages} {1114} (\bibinfo {year}
		{1972})}\BibitemShut {NoStop}%
	\bibitem [{\citenamefont {Boyer}\ and\ \citenamefont
		{Lindquist}(1967)}]{Boyer:1966qh}%
	\BibitemOpen
	\bibfield  {author} {\bibinfo {author} {\bibfnamefont {R.~H.}\ \bibnamefont
			{Boyer}}\ and\ \bibinfo {author} {\bibfnamefont {R.~W.}\ \bibnamefont
			{Lindquist}},\ }\href {\doibase 10.1063/1.1705193} {\bibfield  {journal}
		{\bibinfo  {journal} {J. Math. Phys.}\ }\textbf {\bibinfo {volume} {8}},\
		\bibinfo {pages} {265} (\bibinfo {year} {1967})}\BibitemShut {NoStop}%
	\bibitem [{\citenamefont {Teukolsky}(1973)}]{Teukolsky:1973ha}%
	\BibitemOpen
	\bibfield  {author} {\bibinfo {author} {\bibfnamefont {S.~A.}\ \bibnamefont
			{Teukolsky}},\ }\href {\doibase 10.1086/152444} {\bibfield  {journal}
		{\bibinfo  {journal} {Astrophys. J.}\ }\textbf {\bibinfo {volume} {185}},\
		\bibinfo {pages} {635} (\bibinfo {year} {1973})}\BibitemShut {NoStop}%
	\bibitem [{\citenamefont {Yang}\ \emph {et~al.}(2012)\citenamefont {Yang},
		\citenamefont {Nichols}, \citenamefont {Zhang}, \citenamefont {Zimmerman},
		\citenamefont {Zhang},\ and\ \citenamefont {Chen}}]{Yang:2012he}%
	\BibitemOpen
	\bibfield  {author} {\bibinfo {author} {\bibfnamefont {H.}~\bibnamefont
			{Yang}}, \bibinfo {author} {\bibfnamefont {D.~A.}\ \bibnamefont {Nichols}},
		\bibinfo {author} {\bibfnamefont {F.}~\bibnamefont {Zhang}}, \bibinfo
		{author} {\bibfnamefont {A.}~\bibnamefont {Zimmerman}}, \bibinfo {author}
		{\bibfnamefont {Z.}~\bibnamefont {Zhang}}, \ and\ \bibinfo {author}
		{\bibfnamefont {Y.}~\bibnamefont {Chen}},\ }\href {\doibase
		10.1103/PhysRevD.86.104006} {\bibfield  {journal} {\bibinfo  {journal} {Phys.
				Rev. D}\ }\textbf {\bibinfo {volume} {86}},\ \bibinfo {pages} {104006}
		(\bibinfo {year} {2012})},\ \Eprint {http://arxiv.org/abs/1207.4253}
	{arXiv:1207.4253 [gr-qc]} \BibitemShut {NoStop}%
	\bibitem [{\citenamefont {Regge}\ and\ \citenamefont
		{Wheeler}(1957)}]{Regge:1957td}%
	\BibitemOpen
	\bibfield  {author} {\bibinfo {author} {\bibfnamefont {T.}~\bibnamefont
			{Regge}}\ and\ \bibinfo {author} {\bibfnamefont {J.~A.}\ \bibnamefont
			{Wheeler}},\ }\href {\doibase 10.1103/PhysRev.108.1063} {\bibfield  {journal}
		{\bibinfo  {journal} {Phys. Rev.}\ }\textbf {\bibinfo {volume} {108}},\
		\bibinfo {pages} {1063} (\bibinfo {year} {1957})}\BibitemShut {NoStop}%
	\bibitem [{\citenamefont {Zerilli}(1970)}]{Zerilli:1970se}%
	\BibitemOpen
	\bibfield  {author} {\bibinfo {author} {\bibfnamefont {F.~J.}\ \bibnamefont
			{Zerilli}},\ }\href {\doibase 10.1103/PhysRevLett.24.737} {\bibfield
		{journal} {\bibinfo  {journal} {Phys. Rev. Lett.}\ }\textbf {\bibinfo
			{volume} {24}},\ \bibinfo {pages} {737} (\bibinfo {year} {1970})}\BibitemShut
	{NoStop}%
	\bibitem [{\citenamefont {Tang}\ \emph
		{et~al.}(2025{\natexlab{b}})\citenamefont {Tang}, \citenamefont {Franchini},
		\citenamefont {V{\"o}lkel},\ and\ \citenamefont {Berti}}]{Tang:2025qaq}%
	\BibitemOpen
	\bibfield  {author} {\bibinfo {author} {\bibfnamefont {R.}~\bibnamefont
			{Tang}}, \bibinfo {author} {\bibfnamefont {N.}~\bibnamefont {Franchini}},
		\bibinfo {author} {\bibfnamefont {S.~H.}\ \bibnamefont {V{\"o}lkel}}, \ and\
		\bibinfo {author} {\bibfnamefont {E.}~\bibnamefont {Berti}},\ }\href@noop {}
	{\  (\bibinfo {year} {2025}{\natexlab{b}})},\ \Eprint
	{http://arxiv.org/abs/2512.17786} {arXiv:2512.17786 [gr-qc]} \BibitemShut
	{NoStop}%
	\bibitem [{\citenamefont {Seidel}\ and\ \citenamefont
		{Iyer}(1990)}]{Seidel:1989bp}%
	\BibitemOpen
	\bibfield  {author} {\bibinfo {author} {\bibfnamefont {E.}~\bibnamefont
			{Seidel}}\ and\ \bibinfo {author} {\bibfnamefont {S.}~\bibnamefont {Iyer}},\
	}\href {\doibase 10.1103/PhysRevD.41.374} {\bibfield  {journal} {\bibinfo
			{journal} {Phys. Rev. D}\ }\textbf {\bibinfo {volume} {41}},\ \bibinfo
		{pages} {374} (\bibinfo {year} {1990})}\BibitemShut {NoStop}%
	\bibitem [{\citenamefont {Chandrasekhar}\ and\ \citenamefont
		{Detweiler}(1975)}]{Chandrasekhar:1975zz}%
	\BibitemOpen
	\bibfield  {author} {\bibinfo {author} {\bibfnamefont {S.}~\bibnamefont
			{Chandrasekhar}}\ and\ \bibinfo {author} {\bibfnamefont {S.~L.}\ \bibnamefont
			{Detweiler}},\ }\href {\doibase 10.1098/rspa.1975.0130} {\bibfield  {journal}
		{\bibinfo  {journal} {Proc. Roy. Soc. Lond. A}\ }\textbf {\bibinfo {volume}
			{345}},\ \bibinfo {pages} {145} (\bibinfo {year} {1975})}\BibitemShut
	{NoStop}%
	\bibitem [{\citenamefont {Chandrasekhar}(1976)}]{10.1098/rspa.1976.0022}%
	\BibitemOpen
	\bibfield  {author} {\bibinfo {author} {\bibfnamefont {S.}~\bibnamefont
			{Chandrasekhar}},\ }\href {\doibase 10.1098/rspa.1976.0022} {\bibfield
		{journal} {\bibinfo  {journal} {Proceedings of the Royal Society of London.
				A. Mathematical and Physical Sciences}\ }\textbf {\bibinfo {volume} {348}},\
		\bibinfo {pages} {39} (\bibinfo {year} {1976})}\BibitemShut {NoStop}%
	\bibitem [{\citenamefont {Chandrasekhar}\ and\ \citenamefont
		{Detweiler}(1976)}]{Chandrasekhar:1976zz}%
	\BibitemOpen
	\bibfield  {author} {\bibinfo {author} {\bibfnamefont {S.}~\bibnamefont
			{Chandrasekhar}}\ and\ \bibinfo {author} {\bibfnamefont {S.~L.}\ \bibnamefont
			{Detweiler}},\ }\href {\doibase 10.1098/rspa.1976.0101} {\bibfield  {journal}
		{\bibinfo  {journal} {Proc. Roy. Soc. Lond. A}\ }\textbf {\bibinfo {volume}
			{350}},\ \bibinfo {pages} {165} (\bibinfo {year} {1976})}\BibitemShut
	{NoStop}%
	\bibitem [{\citenamefont {Chandrasekhar}\ and\ \citenamefont
		{Detweiler}(1977)}]{Chandrasekhar:1977kf}%
	\BibitemOpen
	\bibfield  {author} {\bibinfo {author} {\bibfnamefont {S.}~\bibnamefont
			{Chandrasekhar}}\ and\ \bibinfo {author} {\bibfnamefont {S.~L.}\ \bibnamefont
			{Detweiler}},\ }\href {\doibase 10.1098/rspa.1977.0002} {\bibfield  {journal}
		{\bibinfo  {journal} {Proc. Roy. Soc. Lond. A}\ }\textbf {\bibinfo {volume}
			{352}},\ \bibinfo {pages} {325} (\bibinfo {year} {1977})}\BibitemShut
	{NoStop}%
	\bibitem [{\citenamefont {Torres~del Castillo}\ and\ \citenamefont
		{Silva-Ortigoza}(1992)}]{TorresdelCastillo:1992zq}%
	\BibitemOpen
	\bibfield  {author} {\bibinfo {author} {\bibfnamefont {G.~F.}\ \bibnamefont
			{Torres~del Castillo}}\ and\ \bibinfo {author} {\bibfnamefont
			{G.}~\bibnamefont {Silva-Ortigoza}},\ }\href {\doibase
		10.1103/PhysRevD.46.5395} {\bibfield  {journal} {\bibinfo  {journal} {Phys.
				Rev. D}\ }\textbf {\bibinfo {volume} {46}},\ \bibinfo {pages} {5395}
		(\bibinfo {year} {1992})}\BibitemShut {NoStop}%
	\bibitem [{\citenamefont {Arbey}\ and\ \citenamefont
		{Auffinger}(2019)}]{Arbey:2019mbc}%
	\BibitemOpen
	\bibfield  {author} {\bibinfo {author} {\bibfnamefont {A.}~\bibnamefont
			{Arbey}}\ and\ \bibinfo {author} {\bibfnamefont {J.}~\bibnamefont
			{Auffinger}},\ }\href {\doibase 10.1140/epjc/s10052-019-7161-1} {\bibfield
		{journal} {\bibinfo  {journal} {Eur. Phys. J. C}\ }\textbf {\bibinfo {volume}
			{79}},\ \bibinfo {pages} {693} (\bibinfo {year} {2019})},\ \Eprint
	{http://arxiv.org/abs/1905.04268} {arXiv:1905.04268 [gr-qc]} \BibitemShut
	{NoStop}%
	\bibitem [{\citenamefont {Arbey}\ \emph {et~al.}(2025)\citenamefont {Arbey},
		\citenamefont {Calz{\`a}},\ and\ \citenamefont
		{Perez-Gonzalez}}]{Arbey:2025dnc}%
	\BibitemOpen
	\bibfield  {author} {\bibinfo {author} {\bibfnamefont {A.}~\bibnamefont
			{Arbey}}, \bibinfo {author} {\bibfnamefont {M.}~\bibnamefont {Calz{\`a}}}, \
		and\ \bibinfo {author} {\bibfnamefont {Y.~F.}\ \bibnamefont
			{Perez-Gonzalez}},\ }\href {\doibase 10.1016/j.dark.2025.101903} {\bibfield
		{journal} {\bibinfo  {journal} {Phys. Dark Univ.}\ }\textbf {\bibinfo
			{volume} {48}},\ \bibinfo {pages} {101903} (\bibinfo {year} {2025})},\
	\Eprint {http://arxiv.org/abs/2502.17240} {arXiv:2502.17240 [gr-qc]}
	\BibitemShut {NoStop}%
	\bibitem [{\citenamefont {Sasaki}\ and\ \citenamefont
		{Nakamura}(1982{\natexlab{a}})}]{Sasaki:1981kj}%
	\BibitemOpen
	\bibfield  {author} {\bibinfo {author} {\bibfnamefont {M.}~\bibnamefont
			{Sasaki}}\ and\ \bibinfo {author} {\bibfnamefont {T.}~\bibnamefont
			{Nakamura}},\ }\href {\doibase 10.1016/0375-9601(82)90507-2} {\bibfield
		{journal} {\bibinfo  {journal} {Phys. Lett. A}\ }\textbf {\bibinfo {volume}
			{89}},\ \bibinfo {pages} {68} (\bibinfo {year}
		{1982}{\natexlab{a}})}\BibitemShut {NoStop}%
	\bibitem [{\citenamefont {Sasaki}\ and\ \citenamefont
		{Nakamura}(1982{\natexlab{b}})}]{Sasaki:1981sx}%
	\BibitemOpen
	\bibfield  {author} {\bibinfo {author} {\bibfnamefont {M.}~\bibnamefont
			{Sasaki}}\ and\ \bibinfo {author} {\bibfnamefont {T.}~\bibnamefont
			{Nakamura}},\ }\href {\doibase 10.1143/PTP.67.1788} {\bibfield  {journal}
		{\bibinfo  {journal} {Prog. Theor. Phys.}\ }\textbf {\bibinfo {volume}
			{67}},\ \bibinfo {pages} {1788} (\bibinfo {year}
		{1982}{\natexlab{b}})}\BibitemShut {NoStop}%
	\bibitem [{\citenamefont {Sasaki}\ and\ \citenamefont
		{Tagoshi}(2003)}]{Sasaki:2003xr}%
	\BibitemOpen
	\bibfield  {author} {\bibinfo {author} {\bibfnamefont {M.}~\bibnamefont
			{Sasaki}}\ and\ \bibinfo {author} {\bibfnamefont {H.}~\bibnamefont
			{Tagoshi}},\ }\href {\doibase 10.12942/lrr-2003-6} {\bibfield  {journal}
		{\bibinfo  {journal} {Living Rev. Rel.}\ }\textbf {\bibinfo {volume} {6}},\
		\bibinfo {pages} {6} (\bibinfo {year} {2003})},\ \Eprint
	{http://arxiv.org/abs/gr-qc/0306120} {arXiv:gr-qc/0306120} \BibitemShut
	{NoStop}%
	\bibitem [{\citenamefont {Hughes}(2000)}]{Hughes:2000pf}%
	\BibitemOpen
	\bibfield  {author} {\bibinfo {author} {\bibfnamefont {S.~A.}\ \bibnamefont
			{Hughes}},\ }\href {\doibase 10.1103/PhysRevD.62.044029} {\bibfield
		{journal} {\bibinfo  {journal} {Phys. Rev. D}\ }\textbf {\bibinfo {volume}
			{62}},\ \bibinfo {pages} {044029} (\bibinfo {year} {2000})},\ \bibinfo {note}
	{[Erratum: Phys.Rev.D 67, 089902 (2003)]},\ \Eprint
	{http://arxiv.org/abs/gr-qc/0002043} {arXiv:gr-qc/0002043} \BibitemShut
	{NoStop}%
	\bibitem [{\citenamefont {Lo}(2024)}]{Lo:2023fvv}%
	\BibitemOpen
	\bibfield  {author} {\bibinfo {author} {\bibfnamefont {R.~K.~L.}\
			\bibnamefont {Lo}},\ }\href {\doibase 10.1103/PhysRevD.110.124070} {\bibfield
		{journal} {\bibinfo  {journal} {Phys. Rev. D}\ }\textbf {\bibinfo {volume}
			{110}},\ \bibinfo {pages} {124070} (\bibinfo {year} {2024})},\ \Eprint
	{http://arxiv.org/abs/2306.16469} {arXiv:2306.16469 [gr-qc]} \BibitemShut
	{NoStop}%
	\bibitem [{\citenamefont {Pound}\ and\ \citenamefont
		{Wardell}(2021)}]{Pound:2021qin}%
	\BibitemOpen
	\bibfield  {author} {\bibinfo {author} {\bibfnamefont {A.}~\bibnamefont
			{Pound}}\ and\ \bibinfo {author} {\bibfnamefont {B.}~\bibnamefont
			{Wardell}},\ }\href {\doibase 10.1007/978-981-15-4702-7\_38-1} {\  (\bibinfo
		{year} {2021}),\ 10.1007/978-981-15-4702-7\_38-1},\ \Eprint
	{http://arxiv.org/abs/2101.04592} {arXiv:2101.04592 [gr-qc]} \BibitemShut
	{NoStop}%
	\bibitem [{\citenamefont {Konoplya}(2003)}]{Konoplya:2003ii}%
	\BibitemOpen
	\bibfield  {author} {\bibinfo {author} {\bibfnamefont {R.~A.}\ \bibnamefont
			{Konoplya}},\ }\href {\doibase 10.1103/PhysRevD.68.024018} {\bibfield
		{journal} {\bibinfo  {journal} {Phys. Rev. D}\ }\textbf {\bibinfo {volume}
			{68}},\ \bibinfo {pages} {024018} (\bibinfo {year} {2003})},\ \Eprint
	{http://arxiv.org/abs/gr-qc/0303052} {arXiv:gr-qc/0303052} \BibitemShut
	{NoStop}%
	\bibitem [{\citenamefont {Matyjasek}\ and\ \citenamefont
		{Opala}(2017)}]{Matyjasek:2017psv}%
	\BibitemOpen
	\bibfield  {author} {\bibinfo {author} {\bibfnamefont {J.}~\bibnamefont
			{Matyjasek}}\ and\ \bibinfo {author} {\bibfnamefont {M.}~\bibnamefont
			{Opala}},\ }\href {\doibase 10.1103/PhysRevD.96.024011} {\bibfield  {journal}
		{\bibinfo  {journal} {Phys. Rev. D}\ }\textbf {\bibinfo {volume} {96}},\
		\bibinfo {pages} {024011} (\bibinfo {year} {2017})},\ \Eprint
	{http://arxiv.org/abs/1704.00361} {arXiv:1704.00361 [gr-qc]} \BibitemShut
	{NoStop}%
	\bibitem [{\citenamefont {Konoplya}\ and\ \citenamefont
		{Zhidenko}(2023)}]{Konoplya:2023moy}%
	\BibitemOpen
	\bibfield  {author} {\bibinfo {author} {\bibfnamefont {R.~A.}\ \bibnamefont
			{Konoplya}}\ and\ \bibinfo {author} {\bibfnamefont {A.}~\bibnamefont
			{Zhidenko}},\ }\href {\doibase 10.1088/1361-6382/ad0a52} {\bibfield
		{journal} {\bibinfo  {journal} {Class. Quant. Grav.}\ }\textbf {\bibinfo
			{volume} {40}},\ \bibinfo {pages} {245005} (\bibinfo {year} {2023})},\
	\Eprint {http://arxiv.org/abs/2309.02560} {arXiv:2309.02560 [gr-qc]}
	\BibitemShut {NoStop}%
	\bibitem [{\citenamefont {Stein}(2019)}]{Stein:2019mop}%
	\BibitemOpen
	\bibfield  {author} {\bibinfo {author} {\bibfnamefont {L.~C.}\ \bibnamefont
			{Stein}},\ }\href {\doibase 10.21105/joss.01683} {\bibfield  {journal}
		{\bibinfo  {journal} {J. Open Source Softw.}\ }\textbf {\bibinfo {volume}
			{4}},\ \bibinfo {pages} {1683} (\bibinfo {year} {2019})},\ \Eprint
	{http://arxiv.org/abs/1908.10377} {arXiv:1908.10377 [gr-qc]} \BibitemShut
	{NoStop}%
	\bibitem [{\citenamefont {Hussain}(2024)}]{Asad:2024kqm}%
	\BibitemOpen
	\bibfield  {author} {\bibinfo {author} {\bibfnamefont {A.}~\bibnamefont
			{Hussain}},\ }\href@noop {} {\enquote {\bibinfo {title}
			{Potatoasad/{K}err{Q}uasinormal{M}odes.jl: Initial release},}\ }\bibinfo
	{howpublished} {\url{https://doi.org/10.5281/zenodo.14171824}} (\bibinfo
	{year} {2024})\BibitemShut {NoStop}%
	\bibitem{DataAvailabilityStatement}
	No external dataset was used in this study. The results reported in this article were generated using publicly available code cited in the manuscript. No separate raw dataset was generated or analyzed in this study.
	\bibitem [{\citenamefont {Motohashi}(2025)}]{Motohashi:2024fwt}%
	\BibitemOpen
	\bibfield  {author} {\bibinfo {author} {\bibfnamefont {H.}~\bibnamefont
			{Motohashi}},\ }\href {\doibase 10.1103/PhysRevLett.134.141401} {\bibfield
		{journal} {\bibinfo  {journal} {Phys. Rev. Lett.}\ }\textbf {\bibinfo
			{volume} {134}},\ \bibinfo {pages} {141401} (\bibinfo {year} {2025})},\
	\Eprint {http://arxiv.org/abs/2407.15191} {arXiv:2407.15191 [gr-qc]}
	\BibitemShut {NoStop}%
	\bibitem [{\citenamefont {Pombo}\ and\ \citenamefont
		{Pizzuti}(2025)}]{Pombo:2025urp}%
	\BibitemOpen
	\bibfield  {author} {\bibinfo {author} {\bibfnamefont {A.~M.}\ \bibnamefont
			{Pombo}}\ and\ \bibinfo {author} {\bibfnamefont {L.}~\bibnamefont
			{Pizzuti}},\ }\href@noop {} {\  (\bibinfo {year} {2025})},\ \Eprint
	{http://arxiv.org/abs/2511.15796} {arXiv:2511.15796 [gr-qc]} \BibitemShut
	{NoStop}%
	\bibitem [{\citenamefont {Teukolsky}\ and\ \citenamefont
		{Press}(1974)}]{Teukolsky:1974yv}%
	\BibitemOpen
	\bibfield  {author} {\bibinfo {author} {\bibfnamefont {S.~A.}\ \bibnamefont
			{Teukolsky}}\ and\ \bibinfo {author} {\bibfnamefont {W.~H.}\ \bibnamefont
			{Press}},\ }\href {\doibase 10.1086/153180} {\bibfield  {journal} {\bibinfo
			{journal} {Astrophys. J.}\ }\textbf {\bibinfo {volume} {193}},\ \bibinfo
		{pages} {443} (\bibinfo {year} {1974})}\BibitemShut {NoStop}%
	\bibitem [{\citenamefont {Brito}\ \emph {et~al.}(2015)\citenamefont {Brito},
		\citenamefont {Cardoso},\ and\ \citenamefont {Pani}}]{Brito:2015oca}%
	\BibitemOpen
	\bibfield  {author} {\bibinfo {author} {\bibfnamefont {R.}~\bibnamefont
			{Brito}}, \bibinfo {author} {\bibfnamefont {V.}~\bibnamefont {Cardoso}}, \
		and\ \bibinfo {author} {\bibfnamefont {P.}~\bibnamefont {Pani}},\ }\href
	{\doibase 10.1007/978-3-319-19000-6} {\bibfield  {journal} {\bibinfo
			{journal} {Lect. Notes Phys.}\ }\textbf {\bibinfo {volume} {906}},\ \bibinfo
		{pages} {pp.1} (\bibinfo {year} {2015})},\ \Eprint
	{http://arxiv.org/abs/1501.06570} {arXiv:1501.06570 [gr-qc]} \BibitemShut
	{NoStop}%
	\bibitem [{BHP()}]{BHPToolkit}%
	\BibitemOpen
	\href@noop {} {\enquote {\bibinfo {title} {{Black Hole Perturbation
					Toolkit}},}\ }\bibinfo {howpublished}
	{(\href{http://bhptoolkit.org/}{bhptoolkit.org})}\BibitemShut {NoStop}%
	\bibitem [{\citenamefont {Lo}\ \emph {et~al.}(2025)\citenamefont {Lo},
		\citenamefont {Sabani},\ and\ \citenamefont {Cardoso}}]{Lo:2025njp}%
	\BibitemOpen
	\bibfield  {author} {\bibinfo {author} {\bibfnamefont {R.~K.~L.}\
			\bibnamefont {Lo}}, \bibinfo {author} {\bibfnamefont {L.}~\bibnamefont
			{Sabani}}, \ and\ \bibinfo {author} {\bibfnamefont {V.}~\bibnamefont
			{Cardoso}},\ }\href {\doibase 10.1103/PhysRevD.111.124002} {\bibfield
		{journal} {\bibinfo  {journal} {Phys. Rev. D}\ }\textbf {\bibinfo {volume}
			{111}},\ \bibinfo {pages} {124002} (\bibinfo {year} {2025})},\ \Eprint
	{http://arxiv.org/abs/2504.00084} {arXiv:2504.00084 [gr-qc]} \BibitemShut
	{NoStop}%
	\bibitem [{\citenamefont {Li}\ \emph {et~al.}(2021)\citenamefont {Li},
		\citenamefont {Lee}, \citenamefont {Guo},\ and\ \citenamefont
		{Chen}}]{Li:2021zct}%
	\BibitemOpen
	\bibfield  {author} {\bibinfo {author} {\bibfnamefont {P.-C.}\ \bibnamefont
			{Li}}, \bibinfo {author} {\bibfnamefont {T.-C.}\ \bibnamefont {Lee}},
		\bibinfo {author} {\bibfnamefont {M.}~\bibnamefont {Guo}}, \ and\ \bibinfo
		{author} {\bibfnamefont {B.}~\bibnamefont {Chen}},\ }\href {\doibase
		10.1103/PhysRevD.104.084044} {\bibfield  {journal} {\bibinfo  {journal}
			{Phys. Rev. D}\ }\textbf {\bibinfo {volume} {104}},\ \bibinfo {pages}
		{084044} (\bibinfo {year} {2021})},\ \Eprint
	{http://arxiv.org/abs/2105.14268} {arXiv:2105.14268 [gr-qc]} \BibitemShut
	{NoStop}%
	\bibitem [{\citenamefont {Davey}\ \emph {et~al.}(2022)\citenamefont {Davey},
		\citenamefont {Dias}, \citenamefont {Rodgers},\ and\ \citenamefont
		{Santos}}]{Davey:2022vyx}%
	\BibitemOpen
	\bibfield  {author} {\bibinfo {author} {\bibfnamefont {A.}~\bibnamefont
			{Davey}}, \bibinfo {author} {\bibfnamefont {O.~J.~C.}\ \bibnamefont {Dias}},
		\bibinfo {author} {\bibfnamefont {P.}~\bibnamefont {Rodgers}}, \ and\
		\bibinfo {author} {\bibfnamefont {J.~E.}\ \bibnamefont {Santos}},\ }\href
	{\doibase 10.1007/JHEP07(2022)086} {\bibfield  {journal} {\bibinfo  {journal}
			{JHEP}\ }\textbf {\bibinfo {volume} {07}},\ \bibinfo {pages} {086} (\bibinfo
		{year} {2022})},\ \Eprint {http://arxiv.org/abs/2203.13830} {arXiv:2203.13830
		[gr-qc]} \BibitemShut {NoStop}%
	\bibitem [{\citenamefont {Motohashi}\ and\ \citenamefont
		{Noda}(2021)}]{Motohashi:2021zyv}%
	\BibitemOpen
	\bibfield  {author} {\bibinfo {author} {\bibfnamefont {H.}~\bibnamefont
			{Motohashi}}\ and\ \bibinfo {author} {\bibfnamefont {S.}~\bibnamefont
			{Noda}},\ }\href {\doibase 10.1093/ptep/ptac020} {\bibfield  {journal}
		{\bibinfo  {journal} {PTEP}\ }\textbf {\bibinfo {volume} {2021}},\ \bibinfo
		{pages} {083E03} (\bibinfo {year} {2021})},\ \Eprint
	{http://arxiv.org/abs/2103.10802} {arXiv:2103.10802 [gr-qc]} \BibitemShut
	{NoStop}%
	\bibitem [{\citenamefont {Bredberg}\ \emph {et~al.}(2010)\citenamefont
		{Bredberg}, \citenamefont {Hartman}, \citenamefont {Song},\ and\
		\citenamefont {Strominger}}]{Bredberg:2009pv}%
	\BibitemOpen
	\bibfield  {author} {\bibinfo {author} {\bibfnamefont {I.}~\bibnamefont
			{Bredberg}}, \bibinfo {author} {\bibfnamefont {T.}~\bibnamefont {Hartman}},
		\bibinfo {author} {\bibfnamefont {W.}~\bibnamefont {Song}}, \ and\ \bibinfo
		{author} {\bibfnamefont {A.}~\bibnamefont {Strominger}},\ }\href {\doibase
		10.1007/JHEP04(2010)019} {\bibfield  {journal} {\bibinfo  {journal} {JHEP}\
		}\textbf {\bibinfo {volume} {04}},\ \bibinfo {pages} {019} (\bibinfo {year}
		{2010})},\ \Eprint {http://arxiv.org/abs/0907.3477} {arXiv:0907.3477
		[hep-th]} \BibitemShut {NoStop}%
\end{thebibliography}
%

\end{document}